\documentclass[11pt,a4paper]{article}
\usepackage{amsmath,amsfonts,amssymb,amsthm,amstext,amscd,array,bbold}
\usepackage{fancyhdr}
\usepackage[dvips]{graphicx}
\usepackage{makeidx}

\newcommand{\bra}[1]{ \langle #1 |}
\newcommand{\ket}[1]{ | #1 \rangle}

\newcommand{\kket}[1]{ | #1 \rangle\!\rangle}

\newcommand{\braket}[2]{\langle #1 | #2 \rangle}
\newcommand{\twoo}{\mbox{$\substack{\circ\\\circ}$}}
\newcommand{\stc}{\stackrel{\scriptstyle\star}{\scriptstyle ,}}

\newcommand{\weyl}[1]{\twoo\, #1 \,\twoo}
\ifx\ll\undefined
  \newcommand{\ll}{\langle\langle}
\else
  \renewcommand{\ll}{\langle\langle}
\fi

\ifx\implies\undefined
  \newcommand{\implies}{\Longrightarrow}
\else
  \renewcommand{\implies}{\Longrightarrow}
\fi

\newcommand{\pref}[1]{(\ref{#1})}

\newcommand{\id}{\textrm{id}}

\def\tri{\triangleright}

\newcommand{\cD}{{\mathcal D}}

\newcommand{\mbf}[1]{{\boldsymbol {#1} }}
\def\ii{{\,{\rm i}\,}}
\def\dd{{\rm d}}

\newcommand{\eq}{\begin{equation}}
\newcommand{\eqend}{\end{equation}}
\newcommand{\eqa}{\begin{eqnarray}}
\newcommand{\nonueqa}{\begin{eqnarray*}}
\newcommand{\eqaend}{\end{eqnarray}}
\newcommand{\nonueqaend}{\end{eqnarray*}}

\newcommand{\bma}[1]{\begin{array}{#1}}
\newcommand{\ema}{\end{array}}
\newcommand{\bc}{\begin{center}}
\newcommand{\ec}{\end{center}}

\renewcommand{\thefootnote}{\fnsymbol{footnote}}
\newcommand{\newsection}{\setcounter{equation}{0}\section}

\def\appendix#1{\addtocounter{section}{1}\setcounter{equation}{0}
\renewcommand{\thesection}{\Alph{section}}
\section*{Appendix \thesection\protect\indent \parbox[t]{11.715cm} {#1}}
\addcontentsline{toc}{section}{Appendix \thesection\ \ \ #1} }

\newcommand{\complex}{{\mathbb C}} 
\newcommand{\zed}{{\mathbb Z}} 
\newcommand{\nat}{{\mathbb N}} 
\newcommand{\real}{{\mathbb R}} 

\newcommand{\mat}{{\mathbb M}} 
\def\alg{{\mathcal A}}
\def\hil{{\mathcal H}}
\def\twist{{\mathcal F}}
\def\braid{{\mathcal R}}
\def\Ecal{{\mathcal E}}
\def\Mcal{{\mathcal M}}

\newif\ifold             \oldtrue

\def\e{{\,\rm e}\,}

\def\be{\begin{equation}}
\def\ee{\end{equation}}
\def\bea{\begin{eqnarray}}
\def\eea{\end{eqnarray}}
\def\bd{\begin{displaymath}}
\def\ed{\end{displaymath}}

\newcommand{\beq}{\begin{eqnarray}}
\newcommand{\eeq}{\end{eqnarray}}

\def\={~=~}

\makeatletter
\newdimen\normalarrayskip              
\newdimen\minarrayskip                 
\normalarrayskip\baselineskip
\minarrayskip\jot
\newif\ifold             \oldtrue            
\def\arraymode{\ifold\relax\else\displaystyle\fi} 
\def\@arrayskip{\ifold\baselineskip\z@\lineskip\z@
     \else
     \baselineskip\minarrayskip\lineskip2\minarrayskip\fi}
\def\@arrayclassz{\ifcase \@lastchclass \@acolampacol \or
\@ampacol \or \or \or \@addamp \or
   \@acolampacol \or \@firstampfalse \@acol \fi
\edef\@preamble{\@preamble
  \ifcase \@chnum
     \hfil$\relax\arraymode\@sharp$\hfil
     \or $\relax\arraymode\@sharp$\hfil
     \or \hfil$\relax\arraymode\@sharp$\fi}}
\def\@array[#1]#2{\setbox\@arstrutbox=\hbox{\vrule
     height\arraystretch \ht\strutbox
     depth\arraystretch \dp\strutbox
     width\z@}\@mkpream{#2}\edef\@preamble{\halign \noexpand\@halignto
\bgroup \tabskip\z@ \@arstrut \@preamble \tabskip\z@ \cr}%
\let\@startpbox\@@startpbox \let\@endpbox\@@endpbox
  \if #1t\vtop \else \if#1b\vbox \else \vcenter \fi\fi
  \bgroup \let\par\relax
  \let\@sharp##\let\protect\relax
  \@arrayskip\@preamble}
\makeatother

\begin{document}
\begin{titlepage}
\begin{flushright}

\baselineskip=12pt

HWM--07--42\\
EMPG--07--22\\
\hfill{ }\\
November 2007
\end{flushright}

\begin{center}

\vspace{2cm}

\baselineskip=24pt

{\Large\bf Duality and Braiding in Twisted Quantum Field Theory}

\baselineskip=14pt

\vspace{1cm}

{\bf Mauro Riccardi} and {\bf Richard J. Szabo}
\\[4mm]
{\it Department of Mathematics}\\ and\\ {\it Maxwell Institute for
Mathematical Sciences\\ Heriot-Watt University\\ Colin Maclaurin Building,
  Riccarton, Edinburgh EH14 4AS, U.K.}
\\{\tt M.Riccardi@ma.hw.ac.uk} , {\tt R.J.Szabo@ma.hw.ac.uk}
\\[40mm]

\end{center}

\begin{abstract}

\baselineskip=12pt

We re-examine various issues surrounding the definition of twisted
quantum field theories on flat noncommutative spaces. We propose an
interpretation based on nonlocal commutative field redefinitions which
clarifies previously observed properties such as the formal
equivalence of Green's functions in the noncommutative and commutative
theories, causality, and the absence of UV/IR mixing. We use these
fields to define the functional integral formulation of twisted
quantum field theory. We exploit techniques from braided tensor
algebra to argue that the twisted Fock space states of these free
fields obey conventional statistics. We support our claims with a
detailed analysis of the modifications induced in the presence of
background magnetic fields, which induces additional twists by
magnetic translation operators and alters the effective
noncommutative geometry seen by the twisted quantum fields. When two
such field theories are dual to one another, we demonstrate that only
our braided physical states are covariant under the duality.

\end{abstract}

\end{titlepage}
\setcounter{page}{2}

\newpage

\renewcommand{\thefootnote}{\arabic{footnote}}
\setcounter{footnote}{0}

\newsection{Introduction\label{Intro}}

Twisted quantum field theory is a modification of the traditional
approach to noncommutative field theory~\cite{DougNek1,szabo} aimed at
restoring the symmetries of spacetime which are broken by
noncommutativity. It was originally proposed in~\cite{CKNT1} as a
means for obtaining a canonical action of the Poincar\'e group for
quantum field theories on Moyal spaces, thus promoting twisted
Poincar\'e invariance to a similar level as ordinary Poincar\'e
symmetry in conventional relativistic quantum field theory. This
twisted symmetry was subsequently generalized to a deformation of the
bialgebra of diffeomorphisms for Moyal spaces and used to
systematically construct noncommutative theories of
gravity~\cite{Aschieri:2005yw}. Since these foundational works a
number of attempts have been made to understand the origins and
implications of the twisted symmetries, and to investigate if they
cure some of the complications which arise in noncommutative quantum
field theory such as UV/IR mixing, violations of causality, and
non-unitarity. This has led to a number of debates and different
proposals for how to implement twisted Poincar\'e covariance on the
Green's functions of the quantum field theory. One of the goals of
this paper is an attempt to settle some of these issues by defining
and analysing the quantization of twisted fields in as systematic a
way as possible.

One of the main issues surrounding twisted quantum field theory
concerns the proper definition of the correlation functions, for
instance whether one should take star products when defining
quantum averages of fields at separated spacetime points. Through some
approaches it has been claimed that the S-matrix and Green's functions
of twisted quantum field theory are perturbatively equivalent to their
commutative counterparts (see
e.g.~\cite{Balachandran:2007vx,Fiore:2007vg,Joung:2007qv,Oeckl2000}).
In the following we will provide an alternative interpretation of this
somewhat surprising conclusion. We will show that twisted
noncommutative quantum fields can be reformulated in terms of
commutative quantum fields whose arguments are the coordinates of
nonlocal dipole degrees of freedom. This point of view is not
essentially new and has been effectively employed in
e.g.~\cite{Balachandran:2007vx,Balachandran:2007kv}. The novelty of
our approach is that we systematically use the dipole fields to
define and interpret the quantum field theory. We will see that all
quantum correlation functions of twisted noncommutative fields defined
using star produts can be canonically expressed in terms of those of
the commutative dipole field operators. This restores causality and
explains the observations above, including why the twisted quantum
field theory is free from UV/IR
mixing~\cite{Balachandran:2006pi}. However, these averages involve
nonlocal field redefinitions and the mapping back to observables in
the original spacetime coordinates yields correlators which differ
from that of the undeformed field theory. The intractibility of this
map has been noticed in completely different contexts
in~\cite{Fiore:2007vg,Grosse:2007vr,Zahn:2006wt}.

A general framework which emcompasses quantum field theories with Hopf
algebra symmetries is provided by braided quantum field
theory~\cite{Oeckl:1999zu} which uses combinatorial techniques of
braided categories for computing quantum averages. It enables the
formulation of symmetry relations among correlation functions such as
Ward-Takahashi identities which can be systematically formulated on Moyal
spaces~\cite{Riccardi:2007ku,Sasai:2007me}. We will show that much of
this algebraic machinery can be avoided by exploiting the formulation
in terms of commutative dipole operators. Their causality property
enables us to derive explicitly the functional integral defining the
quantum field theory which is manifestly invariant under the twisted
spacetime symmetries. As it is formulated in terms of the nonlocal
dipole coordinates, the functional integration measure differs from
the usual commutative one that is used in the traditional perturbative
approaches to noncommutative quantum field theory which suffer from
UV/IR mixing, acausality and non-unitarity. This point of view
suggests that twisted quantum fields capture the infrared dynamics of
noncommutative field theory which is dual to the ultraviolet dynamics
of the elementary noncommutative quantum fields.

Another central issue surrounds the physical interpretation of states
in the twisted Fock space for free fields. The debate does not seem to
involve the deformation of the canonical commutation relations of
creation and annihilation operators, which has been recently confirmed
to arise from a noncommutative correspondence principle based on
consistent twistings of the bilinear maps
associated to Poisson and commutator brackets
in~\cite{Aschieri:2007sq}. Indeed, our definition of twisted quantum
fields also leads to the same deformation. The main complication is
the claim that the statistics of particle states change due to the
twisting (see e.g.~\cite{Balachandran:2006pi}). In this paper we will
argue that no such change occurs and twisted Poincar\'e invariance is
compatible with conventional Bose or Fermi statistics. This follows
immediately from the dipole formulation which requires defining
multiparticle states using braided tensor algebra, as
in~\cite{Fiore:2007vg}, which is compatible with the star product and
the deformed coproduct of the twisted Hopf symmetry algebra. We show
explicitly that the twisted states obey ordinary statistics under the
twisted action of the permutation group on the Fock space.

While our argument for conventional statistics is rooted in a deep
algebraic fact, the braiding and twist in the case of Moyal spaces is
so simple that it simply amounts to consistent choices of sign
conventions in momentum dependent phases of physical states. To
further elucidate the validity of these choices, we show that only
these definitions map consistently in the expected way between two
noncommutative field theories which are dual to one another. This
situation occurs, for instance, when the field theory involves charged
fields coupled to a constant background magnetic field $F$. If we
allow the two-form $F$ to be a freely varying parameter, then one
obtains a continuous family of twisted quantum field theories labelled
by $F$ and related to each other by duality transformations. We show
that the effective noncommutative geometry seen by the twisted quantum
fields is modified in this instance. In particular, there is a
``self-dual'' point where only a commutative description of twisted
fields is available (in half the spacetime dimension). When the field
theory is defined on a noncommutative torus, the flux $F$ is quantized
and there is an infinite discrete family of twisted quantum field
theories related to each other by Morita equivalence. We adapt the
standard duality mappings between adjoint fields to our definitions of
twisted quantum fields and show that the twisted Fock space states,
with our convention using braiding, transform covariantly under the
Morita duality. This shows that the twisted particle states with
conventional statistics are the only ones which obey the expected
physical equivalence in the mapping between the dual quantum field
theories. As a byproduct of our construction, we show how twisted
oscillators in background fields are modified by further twists
involving magnetic translation operators and relate our construction
to properties of the renormalizable, duality-covariant noncommutative
quantum field theories.

The outline of the remainder of this paper is as follows. Throughout
this paper we will be mostly concerned with the systematic
construction and analysis of twisted scalar quantum fields themselves,
and not the explicit implementation of the twisted spacetime
symmetries. We shall also frequently compare and contrast our results
with the existing literature for clarity. In Section~\ref{TwistMoyal}
we detail the definition of twisted quantum fields in terms of dipole
operators and describe their main properties. In
Section~\ref{TwistFunct} we construct the functional integral
formulation of the twisted quantum field theory and briefly study the
nonlocal field redefinitions required to map correlation functions in
terms of the local spacetime coordinates. In Section~\ref{BraidedFock}
we study the twisted Fock space, and in Section~\ref{TwistMagnetic} we
describe the modifications to twisted quantum field theory for
charged scalar fields coupled to magnetic fields. In
Section~\ref{sec:fockspaceinthedualtheory} we derive the covariant
transformation law for twisted states on a two-dimensional rational
noncommutative torus, while in Section~\ref{DualityIrrational} the
construction is extended to higher dimensions and irrational
noncommutativity parameters.

\newsection{Twisted Quantum Fields on Moyal Spaces\label{TwistMoyal}}

In this section we will define twisted quantum field theory in a way
which naturally explains how the usual pathologies of noncommutative
field theory, such as UV/IR mixing and nonunitarity in Minkowski
space, are cured by, for example, the elimination of non-planar
diagrams from the perturbation expansion. This elucidates some
previous observations in the
literature~\cite{Balachandran:2006pi}. However, we will also stress
the point that correlation functions of the field theory are
\emph{not} the same as those of ordinary commutative field theory,
contrary to some previous
claims~\cite{Abe:2006ig,Balachandran:2007vx,Fiore:2007vg,
  Joung:2007qv,Tureanu:2006pb}.

\subsection{Dipole Coordinates\label{Dipolecoords}}

The simplest noncommutative space is the Moyal space of dimension $d$
which is described by the associative $*$-algebra
$\alg_\theta=\alg_\theta(\real^d)$ generated by hermitean coordinate
operators $\hat x_i$ with $i=1,\dots,d$ subject to a set of Heisenberg
commutation relations. We extend this algebra by linear derivations
$\hat p^i=-\ii\hat\partial^i\in{\rm Aut}(\alg_\theta)$, $i=1,\dots,d$
to a deformed algebra of differential operators
$\cD_\theta=\cD_\theta(\real^d)$. The generators of $\cD_\theta$ obey
the commutation relations
\bea
\big[\hat x_i\,,\,\hat x_j\big] &=& \ii\theta_{ij} \ ,
\nonumber\\[4pt] 
\left[\hat x_i\,,\,\hat p^j\right] &=& \ii\delta_i{}^j \ ,
\nonumber\\[4pt]  \left[ \hat p^i\,,\,\hat p^j\right] &=& 0 \ ,
\label{Dthetarels}\eea
where $\theta=(\theta_{ij})$ is a constant, positive antisymmetric
$d\times d$ noncommutativity parameter matrix (so that the spacetime
dimension $d$ is even). In this paper we will work for the
most part on the trivial rank one module $\cD_\theta$ over this
algebra, acting on itself by left multiplication.

There is an algebra morphism
\bea
\cD_\theta&\longrightarrow&\cD_0 \nonumber\\
\big(\hat x_i\,,\,\hat p^j\big)&\longmapsto&\big(X_i\,,\,P^j\big)
\label{algmor}\eea
defined by the mapping of hermitean operators
\bea
X_i &\doteq& \hat x_i + \mbox{$\frac{1}{2}$}\,\theta_{ij}\,\hat p^j \
, \nonumber\\[4pt]
P^i&\doteq& \hat p^i \ .
\label{Boppshift}\eea
The algebra $\cD_0$ generated by $\{X_i,P^j\}$ is the standard
canonical commutation relation algebra
\bea
\big[X_i\,,\,X_j\big] &=& 0 \ , \nonumber\\[4pt]
\big[X_i\,,\,P^j\big] &=& \ii\delta_i{}^j \ , \nonumber\\[4pt] 
\big[P^i\,,\,P^j\big] &=& 0 \ .
\label{CCRalgQM}\eea
Using (\ref{Dthetarels}) and the fact that the momentum operators
$\hat p^i$ are derivatives, the new coordinate operators $X_i$ can be
rewritten in the form
\begin{equation}
X_i \= \hat x_i -\mbox{$\frac{\ii}{2}$}\,\theta_{ij}\,\big[
\hat\partial^j\,,\,-\big]\=\hat x_i - \mbox{$\frac{1}{2}$}~ {\rm ad}_{
\hat x_i}\=\mbox{$\frac{1}{2}$}\,\big(\hat x_i + \hat x_i^{(R)}\big) \
,
\end{equation}
where $\hat x_i^{(R)}$ denotes the right action of the coordinate
operators on the algebra of observables $\alg_\theta$. The latter form
of $X_i$ can be easily seen to correspond to commuting coordinate
operators~\cite{Balachandran:2007kv,Balachandran:2006ib}.

The operators $X_i$ have appeared before in many different
contexts. In quantum mechanics the morphism (\ref{algmor}) is called
the Bopp shift~\cite{Bopp1956}. In the context of nonlinear integrable
systems its analog is called a dressing
transformation~\cite{Faddeev1980,Grosse1979,Zam1979}. In this paper we
will call $X_i$ ``dipole coordinates'', as they are commuting but
nonlocal position operators which grow with increasing centre of mass
momentum transverse to their extension, due to their dipole moment
$\ell_i=\frac12\,\theta_{ij}\,p^j$. They may be thought of as
parametrizing the fundamental physical excitations responsible for the
nonlocal interactions of noncommutative field
theory~\cite{DougNek1,szabo}. Indeed, the twisted quantum field theory
defined below mimicks the definitions of noncommutative dipole field
theories~\cite{Bergman:2000cw,Dasgupta:2001zu}. The infrared
dynamics of these dipoles are dual to the ultraviolet dynamics of the
elementary noncommutative quantum fields. The twisted quantum field
theory that we study in this paper isolates this low-energy sector of
noncommutative quantum field theory.

Let us recall the geometrical meaning of the morphism
(\ref{algmor})~\cite{Aschieri:2005yw,Balachandran:2007kv,
  Balachandran:2006ib,Riccardi:2007ku,Szabo:2006wx}. Let
\beq
\xi(x)=\xi_i(x)~\partial^i
\eeq 
be a vector field acting in the scalar representation of the
diffeomorphism group ${\rm Diff}(\real^d)$. We identify the
coordinates $x=(x_i)$ of $\real^d$ as the simultaneous eigenvalues of
the operators $X_i$ and $P^i$ with the differential operators
$-\ii\partial^i=-\ii\partial/\partial x_i$. A twisted diffeomorphism
in $\cD_\theta$ can then be obtained from a standard diffeomorphism
generated by $\xi(x)$ by simply identifying the eigenvalues
$x\in\real^d$ with the commutative dipole operators $X$ themselves to
obtain
\bea
\xi^\theta(\hat x)&\doteq&\xi_i(X)~\partial^i\=
\xi_i\big(\hat x - \mbox{$\frac{\ii}{2}$}\,\theta\cdot\hat\partial\,
\big)~\hat\partial^i \nonumber\\[4pt]& =& \xi_i(\hat x)~\hat\partial^i+
 \sum_{n=1}^\infty\,\left(-\frac{\ii}{2}\right)^n\,\frac{1}{n!}\,
\theta_{i_1j_1}\cdots\theta_{i_nj_n}\,\big(\hat\partial^{i_1}
\cdots\hat\partial^{i_n}\xi_i(\hat x)\big)~\hat\partial^{j_1}
\cdots\hat\partial^{j_n}~\hat\partial^i \ .
\label{defvecfields}\eea
This formula agrees with the standard
expression~\cite{Aschieri:2005yw,Szabo:2006wx} for the action of
twisted diffeomorphisms on $\alg_\theta$ as $(\xi^\theta\tri f)(\hat
x)\doteq\xi^\theta(\hat x)\,f(\hat x)=\xi(f)$.

This definition ensures that the symmetry generators $\xi^\theta$ act
covariantly on the algebra $\alg_\theta$. For any pair of
functions $f,g\in\alg_0$ one has the twisted Leibniz rule
\beq
\xi^\theta\triangleright(f\star g)=\mu_\theta\big(\Delta_\theta(
\xi^\theta)\triangleright(f\otimes g)\big) \ ,
\label{twistedLeibniz}\eeq
where
\beq
f\star g\=\mu_\theta(f\otimes g)\=\big(\mathcal{F}^{(1)}\tri f\big)\,
\big(\twist^{(2)}\tri g\big)
\label{starprod}\eeq
is the noncommutative star product on $\alg_0$ with
\beq
\twist\=\twist^{(1)}\otimes\twist^{(2)}\=\exp\big(-\mbox{$\frac\ii2$}\,
\theta_{ij}\,P^i\otimes P^j\,\big)
\label{twistelt}\eeq
an abelian Drinfeld twist associated to the algebra
$\cD_\theta$. Here we have used a Sweedler notation with a suppressed
summation index, and the twisted coproduct is given by
\beq
\Delta_\theta\big(\xi^\theta\big)=\twist^{-1}\,\big(\xi^\theta\otimes
\id_{\alg_\theta}+\id_{\alg_\theta}\otimes\xi^\theta\big)\,\twist \ .
\label{defcoprod}\eeq

The usual observables of noncommutative field theory are defined by
substituting the operators $\hat x$ for the arguments $x$ of
functional expressions for commutative observables
$f\in\alg_0$~\cite{DougNek1,szabo}. With a suitable ordering this
defines the Weyl transform
$W:\alg_0\to\alg_\theta$. In what follows we will be interested in the
subalgebra of $\alg_\theta$ obtained by substituting instead the
dipole coordinates $X$, according to prescription stated above. These
are known as twisted fields and are the ones which transform
covariantly with respect to the action of the deformed vector fields
(\ref{defvecfields}) (when expressed in dipole coordinates). Since the
dipole operators are (linear) combinations of the coordinates and
momenta, they yield nonlocal field redefinitions containing an
infinite number of derivatives in terms of the original variables
$x$. Thus even though the new coordinates $X_i$ commute, their
insertion leads to an inherent nonlocality characteristic of
noncommutative field theory.

\subsection{Free Twisted Quantum Fields\label{FreeDipole}}

Consider a free real scalar quantum field $\check\phi(x)$ of mass
$m>0$ in second quantization on ordinary Minkowski spacetime in $d$
dimensions. We denote the simultaneous eigenvalues of the momentum
operators $P^i$ by $p=(p^i)=(p^0,\mbf p)$, where $\mbf
p\in\real^{d-1}$ and the (upper) mass-shell relation gives
\beq
p^0 =\sqrt{m^2 + \mbf p^2} \ .
\eeq
The mode expansion of the relativistic field is given by
\begin{equation}
\check\phi(x) = \int\, \frac{\dd^{d-1}\mbf p}{2p^0}~
\left(\check a(p) ~\e^{-\ii p\cdot x} + \check a^\dag(p) ~
\e^{\ii p\cdot x}\right) \ ,
\label{modeexpfield}\end{equation}
which realizes it as an operator-valued tempered distribution in terms
of the standard representation of the canonical commutation relation
algebra on the bosonic Fock space $\hil$. The creation and
annihilation operators obey
\begin{equation}
\begin{array}{rcl}
\check a(p) \,\check a(q) & = & \check a(q) \,\check a(p) \ , \\[4pt]
\check a^\dag(p)\, \check a^\dag(q) & = & \check a^\dag(q) \,\check
a^\dag(p) \ , \\[4pt]
\check a(p) \,\check a^\dag(q) & = & \check a^\dag(q) \,\check a(p) +
2p^0 \; \delta^{d}(p-q) \ .
\end{array}
\label{CCRalg}\end{equation}

Let us now substitute the commuting dipole coordinate operators $X$ in
place of their eigenvalues $x$. To make this {\em substitution}
unambiguous, we define it in terms of the symmetrically ordered
parity operator $\weyl{\delta^d(\hat A\,)} \doteq \int
\,\frac{\dd^dp}{(2\pi)^d} ~\exp{(\ii p\cdot\hat A\,)}$, so that for
example the Weyl transform $W:\alg_0\to\alg_\theta$ of a function
$f\in\alg_0$ is given by
\bea
W\big[f(x)\big]\=
f(\hat x)&\doteq&\int\, \dd^d\xi~ f(\xi)~\weyl{\delta^d(\xi-\hat x)}
\nonumber\\[4pt] &=& 
\int \,\frac{\dd^dp}{(2\pi)^d}~ \int \,\dd^d\xi~ f(\xi)~\exp{
\big(\ii p\cdot(\xi-\hat x)\big)} \ .
\label{WeyltransfMoyal}\eea
This definition leads directly to the Moyal product
\beq
f(\hat x)\,g(\hat x)=(f\star g)(\hat x) \ .
\label{WeylMoyalprod}\eeq
For the positive energy component of the quantum field $\check\phi$ at
momentum $p$, we then have
\bea
\check\phi_p^+(X)&=& \check\phi_p^+\big(\hat x +\mbox{$\frac{1}{2}$}\,
\theta\cdot\hat p\big) \nonumber\\[4pt] &=& 
\int\, \dd^d\xi~\e^{-\ii p\cdot\xi}~
\check a(p)~\weyl{\delta^d\big(\xi-\hat x -\mbox{$\frac{1}{2}$}\,
\theta\cdot\hat p\big)} \= \check
a(p)~\e^{-\frac{\ii}{2}\,p\cdot\theta\cdot\hat p}~\e^{-\ii p\cdot\hat
  x} \ .
\label{eqn:twistedcomponent}
\eea
This realizes the twisted quantum field $\check\phi(X)$ as an operator
on $\cD_\theta\otimes\hil$.

Alternatively, we can represent $\check\phi(X)$ as an operator on
$\alg_\theta\otimes\hil$ via the Weyl transform by defining a set of
{\em twisted oscillators} (see for
example~\cite{Aschieri:2007sq,Balachandran:2006pi,
  Fiore:2007vg,Grosse:2007vr}), i.e. by defining a new set of creation
and annihilation operators $\check a^{\phantom{\dagger}}_\theta(p),\check
a^\dagger_\theta(p)$ in terms of the original
ones. Eq.~\pref{eqn:twistedcomponent} suggests the definition
\begin{equation}
\begin{array}{lcl}
\check a_\theta(p) & \doteq & \check a(p)
~\e^{-\frac{\ii}{2}\,p\cdot\theta\cdot\check P} \ , \\[4pt]
\check a_\theta^\dag(p) & \doteq & \check a^\dag(p)~
\e^{\frac{\ii}{2}\,p\cdot\theta\cdot\check{P}}
\end{array}
\label{def:twistedoscill}
\end{equation}
where
\beq
\check P^i \doteq \int\, \frac{\dd^{d-1}\mbf p}{2p^0}~
p^i~ \check a^\dag(p)\, \check a(p)
\label{enmomop}\eeq
is the momentum operator on the Fock space $\hil$. It can be
easily shown that the ordering of the oscillators $\check a(p)$ and
exponentials in (\ref{def:twistedoscill}) is immaterial, and that in
the definition of the momentum operator (\ref{enmomop}) we could as
well have used the $\check a_\theta(p)$ oscillators instead, due to
antisymmetry of the tensor~$\theta_{ij}$.

The new oscillators generate the twisted canonical commutation
relation algebra
\begin{equation}
\begin{array}{lcl}
\check a_\theta(p) \,\check a_\theta(q) & = &  \check a_\theta(q)\,
\check a_\theta(p)~\e^{\ii p\cdot\theta\cdot q} \ , \\[4pt]
\check a_\theta^\dag(p) \,\check a_\theta^\dag(q) & = & \check
a_\theta^\dag(q) \,\check a_\theta^\dag(p)~\e^{\ii p\cdot\theta\cdot
  q} \ , \\[4pt]
\check a_\theta(p) \,\check a_\theta^\dag(q) & = & \check
a_\theta^\dag(q) \,\check a_\theta(p)~\e^{-\ii p\cdot\theta\cdot q} +
2p^0 \; \delta^d(p-q) \ .
\end{array}
\label{twistcommrels}\end{equation}
The field operators corresponding to these twisted oscillators are
given by
\begin{equation}
\check\phi^\theta(x) = \int\, \frac{\dd^{d-1}\mbf p}{2p^0}~
\left(\check a_\theta(p)~\e^{-\ii p\cdot x} + \check a_\theta^\dag(p)~
\e^{\ii p\cdot x}\right) \ .
\label{twistscalarfield}\end{equation}
They can be used to canonically generate the twisted quantum fields
above through their Weyl transform
\begin{equation}
\check\phi(X) = \check\phi^\theta(\hat x) \ .
\label{eqn:shifteqtwist}
\end{equation}

Using the twisted commutation relations (\ref{twistcommrels}), we can
compute the star commutator of the scalar field
(\ref{twistscalarfield}) at separated points. One has
\begin{equation}
\begin{split}
\big[ &\check\phi^\theta(x)\stc\check\phi^\theta(y)\big] 
\\ & \quad\quad \= 
\int\, \frac{\dd^{d-1}\mbf p}{2p^0}~ \int\, \frac{\dd^{d-1}\mbf
  q}{2q^0}~
\bigg[ \Big(\check a_\theta(p)\;\check a_\theta(q)~\e^{-\ii p\cdot
  x}\star \e^{-\ii q\cdot y}-\check a_\theta(q)\;\check a_\theta(p)~
\e^{-\ii q\cdot y}\star \e^{-\ii p\cdot x}\Big) \\
 &\qquad\qquad\qquad\qquad\qquad\qquad\qquad+ \,
\left(\check a_\theta(p)\;\check a_\theta^\dag(q)~
\e^{-\ii p\cdot x}\star \e^{\ii q\cdot y}-\check a_\theta^\dag(q)\;
\check a_\theta(p)~\e^{\ii q\cdot y}\star \e^{-\ii p\cdot x}\right) \\
 &\qquad\qquad\qquad\qquad\qquad\qquad\qquad+\, 
\left(\check a_\theta^\dag(p)\;\check a_\theta(q)~\e^{\ii p\cdot
     x}\star \e^{-\ii q\cdot y}-\check a_\theta(q)\;\check
   a_\theta^\dag(p)~\e^{-\ii q\cdot y}\star \e^{\ii p\cdot x}\right) \\
 &\qquad\qquad\qquad\qquad\qquad\qquad\qquad+ \,
\left(\check a_\theta^\dag(p)\;\check a_\theta^\dag(q)~
\e^{\ii p\cdot x}\star \e^{\ii q\cdot y}-\check a_\theta^\dag(q)\;
\check a_\theta^\dag(p)~\e^{\ii q\cdot y}\star \e^{\ii p\cdot
  x}\right)\bigg] \ .
\end{split}
\label{eqn:startwistfieldcomm}
\end{equation}
From (\ref{Dthetarels}), (\ref{WeylMoyalprod}) and the
Baker-Campbell-Hausdorff formula one has the identity
\begin{equation}
\e^{\ii p\cdot x}\star \e^{\ii q\cdot y}=\e^{\ii p\cdot x+\ii q\cdot
  y}~\e^{-\frac{\ii}{2}\,p\cdot\theta\cdot q} \ .
\label{elemid}\end{equation}
It follows that every bracketed term in \pref{eqn:startwistfieldcomm}
vanishes but one, leaving the result
\begin{equation}
\big[\check\phi^\theta(x)\,\stc\,\check\phi^\theta(y)\big] = 
\ii\,\int \,\frac{\dd^dp}{p^0} ~ \sin{p\cdot(x-y)}~
\delta\big(p^0-\sqrt{m^2 + \mbf p^2} ~\big) \ .
\label{microcausal}\end{equation}
The integral vanishes for spacelike separated points $x,y$, and hence
the algebra of twisted real scalar fields with the star product is
microcausal, just as in the commutative case. This property is
consistent with their identification (\ref{eqn:shifteqtwist}) in terms
of commutative dipole fields and was also observed
in~\cite{Fiore:2007vg}.

However, as the dipole coordinate $X$ is defined in terms of {\em
  both} coordinates and momenta, the field $\check\phi(X)$ is
expressed in a nonlocal way in terms of the original position variable
$x$. It follows that local observables written in terms of
$\check\phi(X)$ correspond to nonlocal observables in terms of the
fields $\check\phi(x)$ containing an infinite number of
derivatives. This is easily seen by noting that the Green's functions
of the quantum field theory are built from the vacuum expectation
values of products of the fields
\bea
\check\phi\big(X^1\big)\cdots \check\phi\big(X^n\big)&=&
\check\phi^\theta\big(\hat x^1\big)\cdots
\check\phi^\theta\big(\hat x^n\big)\=
W\big[\check\phi^\theta\big(x^1\big)
\star\cdots\star\check\phi^\theta\big(x^n\big) \big] \nonumber \\[4pt]
&=&  W\Big[\,\prod_{a<b}\,\exp\Big(-\frac\ii2\,\frac\partial
{\partial x_i^a}\,\theta_{ij}\,\frac\partial{\partial x_j^b}\,
\Big)\check\phi^\theta\big(x^1\big)\cdots
\check\phi^\theta\big(x^n\big)\,\Big] \ .
\eea
In turn, this implies that there are no non-planar diagrams in the
perturbation expansion of any interacting model built on these fields,
and hence the usual undesirable features of noncommutative quantum
field theory such as UV/IR mixing and non-unitarity in Minkowski
signature do not show up in twisted quantum field theory. There is,
however, a nonlocal correspondence between the correlation functions
of the twisted theory and those of the original untwisted
noncommutative field theory. This renders an alternative
perspective on the problem of the computability of twisted quantum
field theory described in~\cite{Zahn:2006wt}. We describe some aspects
of this correspondence in the next section.

\newsection{Functional Formulation of Twisted Quantum Field
  Theory\label{TwistFunct}}

The definition of the functional integral which defines a quantum
field theory invariant under twisted spacetime symmetries can be given
formally in the context of braided quantum field
theory~\cite{Oeckl:1999zu,Oeckl2000,Sasai:2007me} (see
also~\cite{Grosse:2001pr}). In this section we will give a more
pragmatic definition which avoids the use of this abstract algebraic
machinery by exploiting the formulation in terms of dipole
fields. This quantization of the twisted field theory makes it
qualitatively similar to noncommutative dipole field
theories~\cite{Dasgupta:2001zu}. The mapping of correlators in the
dipole coordinates back to correlators in the original spacetime
coordinates can be presumably achieved using the braided formalism,
though we will not pursue this issue here.

\subsection{Twist Invariant Functional Integral}

To quantize the noncommutative scalar field theory defined in terms of
twisted fields using functional methods, we need to construct a
measure for the path integral which defines the twisted quantum field
theory with the manifest twisted spacetime symmetries. This can be
done by using the formalism of dipole coordinates and mimicking the
standard treatment of the functional integral in the commutative
case. Notice first of all that the field theory with twisted fields
and the star product can be canonically quantized, since the equal
time canonical commutation relations hold. Writing $x=(\mbf x;t)$ with
$\mbf x\in\real^{d-1}$, in the free field case this follows by
applying time derivatives to (\ref{microcausal}).

The equal time canonical commutation relation algebra is
\bea
\big[\check\phi^\theta(\mbf x;t)\stc\check\phi^\theta(\mbf y;t)\big]
&=& 0 \ , \nonumber\\[4pt]
\big[\check\Pi^\theta(\mbf x;t)\stc\check\Pi^\theta(\mbf y;t)\big] &=&
0 \ , \nonumber\\[4pt] 
\big[\check\phi^\theta(\mbf x;t)\stc\check\Pi^\theta(\mbf y;t)\big]
&=& \ii\delta^{d-1}(\mbf x-\mbf y)
\eea
where $\check\Pi^\theta(\mbf x;t)$ is the momentum operator conjugate
to the quantum field $\check\phi^\theta(\mbf x;t)$. One can therefore
apply all the usual machinery of coherent states
$\ket{\phi;t},\ket{\Pi;t}$ defined as the simultaneous eigenstate
functionals of the quantum field operators with
\begin{equation}
\check\phi(\mbf X;t)\ket{\phi;t} ~\doteq~
\phi(\mbf\xi;t)\,\ket{\phi;t}  
\qquad
\mbox{and} \qquad \check\Pi(\mbf X;t)\ket{\Pi;t} ~\doteq~
\Pi(\mbf\xi;t) \,\ket{\Pi;t}
\end{equation} 
where $\mbf\xi\in\real^{d-1}$ are the eigenvalues of the spatial
dipole operators $\mbf X$. They span a functional Hilbert space
$\underline{\hil}$ and have the usual inner products
\bea
\braket{\phi_1;t}{\phi_2;t} &=& \delta[\phi_1-\phi_2] \ ,
\nonumber\\[4pt]
\braket{\Pi_1;t}{\Pi_2;t} &=& \delta[\Pi_1-\Pi_2] \ ,
\nonumber\\[4pt]
\braket{\phi;t}{\Pi;t} &=& \frac{1}{\sqrt{2\pi}}\, \exp{\Big(\ii \,
\int\,\dd^{d-1}\mbf\xi~ \Pi(\mbf\xi;t)\, \phi(\mbf\xi;t)\Big)} \ .
\label{fninnerprods}\eea
We only have to keep track of star products when they appear.

Let us start with the propagator
\begin{equation}
\braket{\,\underline{\Psi}\,_{\rm out}(t_{\rm f})}{\,
\underline{\Psi}\,_{\rm in}(t_{\rm i})}
\label{propagator}\end{equation}
describing the S-matrix element between in and out states. We
partition the time interval $[t_{\rm i},t_{\rm f}]$ into intermediate
times $t_{\rm i} =t_0 < t_1 < \dots < t_N = t_{\rm f}$, and at the end
take the limit $N\to\infty$. We use the completeness relation
\begin{equation}
\id_{\underline{\hil}}
= \int~ \prod_{\mbf\xi\in\real^{d-1}}\, \dd\phi(\mbf\xi;t)~ 
\ket{\phi;t}\bra{\phi;t}
\label{phicomplid}\end{equation}
with a redundant notation to single out the time dependence. Inserting
the identity (\ref{phicomplid}) for each time $t_l$ into the
propagator (\ref{propagator}), we obtain
\begin{equation}
\begin{split}
\braket{\,\underline{\Psi}\,_{\rm out}(t_{\rm f})}{\,
\underline{\Psi}\,_{\rm in}(t_{\rm i})} 
\= & \lim_{N\to\infty}~
\prod_{l=0}^N ~\int~\prod_{\mbf\xi_l\in\real^{d-1}}\,
\dd\phi_l(\mbf\xi_l;t_l)~ \braket{\,
\underline{\Psi}\,_{\rm out}(t_N)}{\phi_N;t_N}\,
\braket{\phi_N;t_N}{\phi_{N-1};t_{N-1}}
\\ &\qquad\qquad \times \, 
\braket{\phi_{N-1};t_{N-1}}{\phi_{N-2};t_{N-2}}
\cdots \braket{\phi_1;t_1}{\phi_0;t_0}\,
\braket{\phi_0;t_0}{\,\underline{\Psi}\,_{\rm in}(t_0)} \ .
\end{split}
\label{propphicompl}\end{equation}

Now we use the fact that the quantum hamiltonian
$H[\check\Pi,\check\phi\,]$ of the scalar field theory generates time
translations with
\begin{equation}
\braket{\phi_2;t_2}{\phi_1;t_1} = \bra{\phi_2;t_1} 
\e^{-\ii(t_2-t_1)\, H[\check\Pi,\check\phi\,]}\ket{\phi_1;t_1} \ .
\end{equation}
Inserting the completeness relation
\begin{equation}
\id_{\underline{\hil}}
= \int~ \prod_{\mbf\xi\in\real^{d-1}}\, \dd\Pi(\mbf\xi;t) ~
\ket{\Pi;t}\bra{\Pi;t}
\end{equation}
at times $t_0,t_1,\dots,t_{N-1}$ into (\ref{propphicompl}) and using
the inner products (\ref{fninnerprods}), we obtain the standard
expression for the phase space path integral
\begin{equation}
\begin{split}
& \braket{\,\underline{\Psi}\,_{\rm out}(t_{\rm f})}{\,
\underline{\Psi}\,_{\rm in}(t_{\rm i})} \\ &
\qquad\qquad \= 
\lim_{N\to\infty}~\int~\prod_{l=0}^N~ \prod_{\mbf\xi_l\in\real^{d-1}}\,
\dd\phi_l(\mbf\xi_l;t_l)~ \prod_{l=0}^{N-1}~
\prod_{\mbf\xi_l\in\real^{d-1}}\,\frac{\dd\Pi_l(\mbf\xi_l;t_l)}
{\sqrt{2\pi}}~\braket{\,\underline{\Psi}\,_{\rm out}(t_N)}{\phi_N;t_N} 
\\ &\qquad\qquad\qquad\qquad\qquad\qquad \times\,
\exp{\Big(\ii\,\sum_{l=0}^{N-1}\, \big(\Pi_l \,(\phi_{l+1}-\phi_l) - 
\delta t_l~ H[\Pi_l,\phi_l]\big)\Big)}\,
\braket{\phi_0;t_0}{\,\underline{\Psi}\,_{\rm in}(t_0)}
\end{split}
\label{phasespacepathint}\end{equation}
with $\delta t_l=t_{l+1}-t_l$. The quantum hamiltonian
$H[\check\Pi,\check\phi\,]$ is the original one defined in terms of
the noncommutative star product, and hence the exponential function in
(\ref{phasespacepathint}) reproduces the usual Boltzmann weight with
the action functional of the untwisted noncommutative field
theory.

The measure in (\ref{phasespacepathint}) is a product of the Lebesgue
measures $\dd\phi(\mbf\xi;t)~\dd\Pi(\mbf\xi;t)$ over the
eigenvalues of the dipole operators $\mbf X$. This defines the twist
invariant functional integration measure. By means of the
identification \pref{eqn:shifteqtwist}, we see that the hamiltonian
written in terms of the field variables $\phi(X)$ is the hamiltonian
of the underlying commutative field theory. However, our S-matrix
elements are not commutative ones but appear to agree with those
computed in~\cite{Bu:2006ha}. Moreover, our functional integration
measure is different from the naive commutative one used in
e.g.~\cite{Tureanu:2006pb}.

\subsection{Coherent State Representation of Dipole Operators}

We will now elucidate the correspondence between the set of
noncommuting coordinates $\hat x$ and the commutative dipole
coordinates $X$, required to match correlation functions defined by
the functional integral (\ref{phasespacepathint}). For this, we
rewrite the standard formalism of coherent state representations
appropriate to the field dependence of twisted quantum field
theory. For clarity we restrict the construction to $d=2$ spacetime
dimensions. First, we pass to complex coordinate operators
\begin{equation}
\begin{array}{rcl}
\hat z & \doteq & \hat x_1 + \ii\hat x_2  \ , \\[4pt]
\hat z^\dagger & \doteq & \hat x_1 - \ii\hat x_2
\end{array}
\end{equation}
obeying the commutation relations
\beq
\big[\hat z\,,\,\hat z^\dagger \big] \= 2\theta_{12}~\doteq~
2\theta \ .
\label{complexcommrels}\eeq
We also define corresponding complex derivations
$\hat\partial,\hat{\bar\partial}$ in such a way that they fulfill the
commutation relations
\begin{equation}
\begin{array}{rcl}
\big[\hat \partial\,,\,\hat z\big] & = & 1 \=
\big[\hat{\bar\partial} \,,\,\hat z^\dagger \big] \ , \\[4pt]
\big[\hat \partial\,,\,\hat z^\dagger \big] & = & 0 \= 
\big[\hat{\bar\partial}\,,\,\hat z\big] \ , \\[4pt]
\big[\hat \partial\,,\,\hat{\bar\partial} \,\big] & = & 0 \ .
\end{array}
\end{equation}
We can then introduce a module over the algebra $\alg_\theta$ spanned
by coherent states which are the normalized eigenstates of the
coordinate operators $\hat z$ given by
\begin{equation}
\ket\zeta \= \exp{\Big(-\frac{|\zeta|}{4\theta}^2\,\Big)}\,
\exp{\Big(\frac{\zeta\,\hat z^\dagger}{2\theta}\Big)}\ket{\zeta=0}
\qquad \mbox{with} \quad \hat z\ket\zeta \= \zeta\,\ket\zeta \ ,
\label{cohstatesdef}\end{equation}
where $\zeta\in\complex$.

We now define the commutative complex dipole coordinates $Z,Z^\dagger$
by
\begin{equation}
\begin{array}{rcl}
Z & \doteq & X_1 + \ii X_2 \= \hat z -\theta\,\hat{\bar\partial} \ ,
\\[4pt]
Z^\dagger & \doteq & X_1 - \ii X_2 \= \hat z^\dagger +
\theta\,\hat\partial
\end{array}
\end{equation}
which obey
\beq
\big[Z\,,\,Z^\dagger \big] =0 \ .
\label{complexdipolerels}\eeq
In order to define the delta-function in these new coordinates, we
need to compute the matrix elements of the plane wave operator
\begin{equation}
\begin{split}
\bra{\zeta'}\exp{(\ii p\cdot X)}\ket{\zeta} \= & \ \bra{\zeta'}
\exp{\big(\mbox{$\frac{\ii}{2}$}\,(p\,Z^\dagger+p^*\,Z)
\big)}\ket{\zeta} \\[4pt]
\= & \ \bra{\zeta'}\exp{\big(\mbox{$\frac{\ii}{2}$}\,\theta\,
(p\,\hat\partial -p^*\,\hat{\bar\partial}\,)\big)}\ket{\zeta}\, 
\exp{\big(\mbox{$\frac{\ii}{2}$}\,(p\,\zeta'^*+p^*\,\zeta)\big)} \ .
\end{split}
\end{equation}
One can straightforwardly derive the identities
\beq
\begin{split}
\bra{\eta}\hat\partial\,^l\ket{\zeta} \= & 
\left(\mbox{$\frac{\eta^*}{2\theta}$}\right)^l\,\braket{\eta}{\zeta} \
,
\\[4pt] 
\bra{\eta}\hat{\bar\partial}\,^l\ket{\zeta} \= & \left(-\mbox{$
\frac{\zeta}{2\theta}$}\right)^l\,\braket{\eta}{\zeta}
\end{split}
\eeq
from which we find
\begin{equation}
\begin{split}
\bra{\zeta'}\exp{(\ii p\cdot X)}\ket{\zeta} \= & \ \bra{\zeta'}
\exp{\big(\mbox{$\frac{\ii}{2}\,\theta\,(p\,\frac{\zeta'^*}{2\theta}
    +p^*\,\frac{\zeta}{2\theta})$}\big)}\ket{\zeta} \,
\exp{\big(\mbox{$\frac{\ii}{2}$}\,(p\,\zeta'^*+p^*\,\zeta)\big)}
\\[4pt]
\= & \exp{\big(\mbox{$\frac{3\ii}{4}$}\,(p\,\zeta'^*+p^*\,
\zeta)\big)} \,\braket{\zeta'}{\zeta} \ .
\end{split}
\label{planewaveX}\end{equation}

We can compare the matrix element (\ref{planewaveX}) with the
analogous one for the standard noncommutative coordinates $\hat
z,\hat{\bar z}$. One finds
\beq
\begin{split}
\bra{\zeta'}\exp{(\ii p\cdot\hat{x})}\ket{\zeta} \= & \ 
\bra{\zeta'}\exp{\big(\mbox{$\frac{\ii}{2}$}\,(p\,\hat z^\dagger+p^*\,
\hat z)\big)}\ket{\zeta} \\[4pt]
\= &  \exp{\big(\mbox{$\frac{\ii}{2}\,(p\,\zeta'^*+p^*\,
\zeta)-\frac{\theta}{4}\,|p|^2$}\big)} \,\braket{\zeta'}{\zeta} \ .
\end{split}
\label{planewavehatx}\eeq
The difference between the two matrix elements (\ref{planewaveX}) and
(\ref{planewavehatx}) emphasizes the commutativity of the dipole
coordinates $X$, as well as the nonlocal relationship between them and
the noncommutative coordinates $\hat x$.

One can also easily derive the matrix elements
\beq
\bra{\zeta'}Z\ket{\zeta} \= \big(\zeta - \mbox{$\frac{1}{2}$}\,
\zeta'^*\big)\,\braket{\zeta'}{\zeta}\qquad \mbox{and} \qquad
\bra{\zeta'}Z^\dagger\ket{\zeta} \= \big(\zeta'^* -
\mbox{$\frac{1}{2}$}\,\zeta\big)\,\braket{\zeta'}{\zeta} \ .
\eeq
From these expressions one can derive a representation of the complex
dipole field operators in the coherent state basis as
\begin{equation}
Z \= \int \,\dd^2\zeta ~\left(\zeta-\mbox{$\frac{1}{2}$}\,
\zeta^*\right)~\ket{\zeta}\bra{\zeta} \qquad \mbox{and} \qquad
Z^\dagger \= \int \,\dd^2\zeta ~\left(\zeta^*-\mbox{$\frac{1}{2}$}\,
\zeta\right)~\ket{\zeta}\bra{\zeta} \ .
\end{equation}
This is to be contrasted with the coherent state representation of the
noncommutative complex coordinate operators, given from the definition
(\ref{cohstatesdef}) by
\begin{equation}
\hat z \= \int \,\dd^2\zeta ~\zeta~ \ket{\zeta}\bra{\zeta} \qquad 
\mbox{and} \qquad \hat z^\dagger \= \int \,\dd^2\zeta ~\zeta^*~
\ket{\zeta}\bra{\zeta} \ .
\end{equation}
The nonanalytic dependence of the dipole coordinate operators is
analogous to that in the definition of Bargmann space in the
quantum mechanics of the lowest Landau Level, where the projection
onto the ground state induces the representation $\zeta^* \mapsto
\partial/\partial\zeta$. This allows one to deal with nonanalytic
potentials in a space of analytic functions, at the expense of
locality~\cite{bargmann,girvinjach}.

\newsection{Braided Tensor Algebra on the Twisted Fock
  Space\label{BraidedFock}}

We have seen that the twisted quantum field theory, although
commutative, has nontrivial features because of its nonlocality in the
original coordinate variables. The perturbation expansion of any
observable, however, has no nonplanar diagrams because of the twist of
the quantum fields. In this section we will explore the physical
meaning of the twisted oscillators $\check a_\theta(p)$ introduced in
(\ref{def:twistedoscill}). We will do this by investigating the
structure of the Fock space $\hil^\theta$ built on the quantum
operators~$\check a_\theta(p)$.

\subsection{Action of the Symmetric Group\label{SymGroupAction}}

The Fock space $\hil$ of the untwisted quantum field theory is graded
by particle number $N\in\nat_0$ as
\beq
\hil=\bigoplus_{N=0}^\infty\,\hil_N \ .
\label{Fockgrading}\eeq
The one-dimensional subspace $\hil_0$ is spanned by the vacuum state
$\ket{\Omega}$ annihilated by the oscillators $\check a(p)$. From the
definition (\ref{def:twistedoscill}) it follows that the vector
$\ket{\Omega}$ is the (unique) vacuum state of the twisted oscillators
$\check a_\theta(p)$. Moreover, since for any momentum $p$ the Fock
space number operator can be written as
\begin{equation}
\check N_p \= \check a^\dag(p) \,\check a(p) \= \check
a_\theta^\dag(p) \, \check a_\theta(p) \ ,
\end{equation}
it follows that any monomial in the twisted oscillators has the naive
particle number. Thus $\hil_0^\theta=\hil_0$ and the grading operator
on the twisted Fock space $\hil^\theta$ is the same as that on
(\ref{Fockgrading}). The one-particle sector $\hil_1^\theta$ is
spanned by the vectors $|p\rangle_\theta\doteq\check
a_\theta^\dag(p)\ket{\Omega}=\check a^\dag(p)\ket{\Omega}$. Thus
$\hil_1^\theta=\hil_1$ and hence single-particle states are momentum
eigenstates which are also unaffected by the twist. This is consistent
with the fact that the twisting does not change the coproducts
(\ref{defcoprod}) of the momentum operators (\ref{enmomop}).

Let us now consider an indecomposable two-particle state in
$\hil_2^\theta$ given by
\begin{equation}
\ket{p}_\theta\otimes\ket{q}_\theta\doteq \check a_\theta^\dag(p) \,
\check a_\theta^\dag(q) \ket{\Omega} \ .
\end{equation}
The permutation of the two particles is determined by a flip operator
$\sigma$ which preserves the grading on $\hil^\theta$. Using the twisted
commutation relations (\ref{twistcommrels}) it is given explicitly by
\bea
\sigma \ket{p}_\theta\otimes\ket{q}_\theta ~\doteq~\ket{q}_\theta
\otimes\ket{p}_\theta&=&\check a_\theta^\dag(q) \,
\check a_\theta^\dag(p) \ket{\Omega} \nonumber\\[4pt] &=& 
\e^{-\ii p\cdot\theta\cdot q}\ 
\check a_\theta^\dag(p)\, \check a_\theta^\dag(q) \ket{\Omega} \= 
\e^{-\ii p\cdot\theta\cdot q}\,\ket{p}_\theta\otimes\ket{q}_\theta \ .
\label{flipdef}\eea
It is easy to see from this equation that the flip operator $\sigma$
fails to be an involution of $\hil^\theta$. To generate a
representation of the permutation group $S_N$ on $\hil^\theta$, we
represent the Drinfeld twist (\ref{twistelt}) as an operator
$\check\twist$ on $\hil^\theta$ by replacing the translation
generators $P^i$ by the second quantized momentum operators
(\ref{enmomop}) acting on $\hil^\theta$. Using this Fock space
representation we define the twisted flip operator
\begin{equation}
\sigma_\theta\doteq \check\twist\circ\sigma\circ\check\twist^{-1} \ .
\label{twistflipdef}\end{equation}
Then $\sigma_\theta$ is an involution of $\hil^\theta$. The
representation of the twist operator $\check\twist$ on $\hil_2^\theta$
extends to arbitrary tensor products of single-particle states where
it satisfies analogous equations.

The meaning of the expression (\ref{twistflipdef}) can be understood
as follows. The flip operator $\sigma$ defines a representation of the
permutation group on the untwisted Fock space $\hil$. After the twist,
it fails to be a representation of the symmetric group on
$\hil^\theta$, until we twist the representation itself, which
restores the Yang-Baxter equations giving the relations of
$S_N$. Applying the twisted flip operator to a two-particle state
yields
\begin{equation}
\sigma_\theta\ket{p}_\theta\otimes\ket{q}_\theta \= \e^{\ii
  p\cdot\theta\cdot q}\,
\check a_\theta^\dag(q)\, \check a_\theta^\dag(p) \ket{\Omega} \=  
\check a_\theta^\dag(p)\, \check a_\theta^\dag(q) \ket{\Omega} \=
\ket{p}_\theta\otimes\ket{q}_\theta \ .
\end{equation}
The relationship between the original state and its image under
transposition is nonlocal in the naive way, because of the appearence
of a differential operator with infinitely many derivatives.

It is interesting to observe that, whilst in the generic case the
operator $\sigma$ is not an involution of the twisted Fock space, by
(\ref{flipdef}) it squares to the identity when the momenta of the
particles belong to an appropriate lattice, as is the case when the
field theory is defined on some torus. Consider for definiteness the
two-dimensional case, and set
\begin{equation}
p^i\=\frac{2\pi}{L_i}\,n_i \qquad \mbox{with} \quad n_i\in\mathbb Z \
, 
\end{equation}
where $L_i$, $i=1,2$ are the sides of a rectangular lattice
$\Gamma\cong\zed^2$. Then the argument of the exponential in
(\ref{flipdef}) is of the form
\begin{equation}
p\cdot\theta\cdot q = 2\pi\,\frac{2\pi\,\theta}{L_1\, L_2}\,
(n_1\, m_2-m_1 \,n_2) \ ,
\end{equation}
and the exponential is unity whenever the quantity
$\frac{2\pi\,\theta}{L_1\, L_2}$ is an integer. When this happens, we
can give the canonical meaning to the exchange of two particles. If
our quantum field theory is defined on a torus whose area is quantized
according to
\begin{equation}
n \cdot\textrm{Area} \= 2\pi\,\theta \qquad \mbox{for} \quad
n\in\mathbb N \ ,
\end{equation}
then there is no need to twist the flip operator as well. In this case
the Fock space $\hil^\theta$ built by means of the twisted creation
and annihilation operators $\check
a_\theta^{\phantom{\dagger}}(p),\check a_\theta^\dagger(p)$ is
bosonic. This simple observation will be elucidated and generalized in
Sections~\ref{sec:fockspaceinthedualtheory}
and~\ref{DualityIrrational}. An analogous fermionic Fock space can be
constructed by starting from untwisted fermionic field operators. 

In the general case, the twisted flip operator $\sigma_\theta$, together
with its natural extensions to higher number multiparticle states,
provides a representation of the permutation group on the twisted Fock
space $\hil^\theta$. It would be interesting to relate this action of
$S_N$ to appropriate generalizations of the known braid group
representations to accomodate our commutation relations, which may be
expressed as
\begin{equation}
\ket{p}_\theta\otimes\ket{q}_\theta \= \e^{\ii p\cdot \theta\cdot
  q}\,\ket{q}_\theta\otimes\ket{p}_\theta \= \check\braid^{-1}
\ket{q}_\theta\otimes\ket{p}_\theta \ .
\label{braidcommrels}\end{equation}
Here
\beq
\braid\=\braid^{(1)}\otimes\braid^{(2)}\=\twist^{2}
\label{braidmatrix}\eeq
is the universal braiding matrix of the triangular Hopf algebra
structure of
$\cD_\theta$~\cite{Aschieri:2005zs,Fiore:2007vg}. Triangularity
implies that the twisted flip operator (\ref{twistflipdef}) can be
written as
\beq
\sigma_\theta\=\check\braid\circ\sigma\=\check\braid^{-1} \ ,
\label{sigmabraid}\eeq
and hence braiding and transposition coincide on the twisted Fock
space $\hil^\theta$.

The commutation relations (\ref{braidcommrels}) uniquely follow from
the twisting of the creation and annihilation operators $\check
a^{\phantom{\dagger}}_\theta(p),\check a^\dagger_\theta(p)$, which is
equivalent to the morphism (\ref{algmor}), and they describe
multiparticle states with ordinary (untwisted) statistics. This result
agrees with those of e.g.~\cite{Bu:2006ha,Fiore:2007vg,Tureanu:2006pb},
but disagrees with results of
e.g.~\cite{Balachandran:2007vx}--\cite{Balachandran:2006pi}
and~\cite{Joung:2007qv} where it was claimed that the states of
$\hil^\theta$ obey twisted statistics. Our formulation is in the
spirit of~\cite{Fiore:1996tr} where conventional Bose and Fermi
statistics are maintained by the twisting of quantum group symmetries,
and of~\cite{Kulish2006} where unitary transformations of operators
such as (\ref{twistflipdef}) are consistently compensated by
transformations of their representations as well (as done
in~\cite{Zam1979}). As we demonstrate below, it is this definition
which follows from a consistent treatment of the star products of
fields that deforms the tensor algebra of $\alg_\theta$ to the braided
tensor algebra corresponding to the triangular structure
$\braid$~\cite{Aschieri:2005zs,Fiore:2007vg}. These deformations are
all required for compatibility with the noncocommutative twisted
coproduct $\Delta_\theta$ and the action of twisted spacetime
symmetries in (\ref{twistedLeibniz}).

\subsection{Braiding of Multiparticle States}

We will now derive a relation between twisted quantum fields and the
definition of twisted multiparticle states of $\hil^\theta$ as {\em
  braided} tensor products of single-particle states. To understand
how this definition arises, we consider the associativity law for the
star product of $N$ fields $f_a(x)$, $a=1,\dots,N$, with respective
Fourier transforms $\tilde f_a(p)$. By iterating the identity
(\ref{elemid}) we compute
\begin{equation}
\begin{split}
f_1(x)\star \cdots\star f_N(x) \= & \prod_{a=1}^N~\int\,\frac{\dd^d
  p_a}{(2\pi)^d}~\tilde f_a(p_a)~\e^{\ii p_1\cdot x}\star 
\cdots\star \e^{\ii p_N\cdot x} \\[4pt]
\= & \prod_{a=1}^N~\int \,\frac{\dd^d p_a}{(2\pi)^d} \ \tilde
f_a(p_a)~\e^{\ii p_a\cdot x}~\exp\Big(
{-\frac{\ii}{2}\,\sum_{a<b}\, p_a\cdot\theta\cdot p_b}\Big) \ .
\end{split}
\end{equation}

When applied to the mode expansion (\ref{modeexpfield}) of an
untwisted quantum field on $\hil$, this relation implies that the
basic multiparticle states of noncommutative quantum field theory in
the basis of momentum eigenstates are given by
\begin{equation}
\begin{split}
\kket{p_1}_\theta\otimes\cdots\otimes\kket{p_N}_\theta
\doteq & \exp\Big(
{-\frac{\ii}{2}\,\sum_{a<b}\, p_a\cdot\theta\cdot p_b}\Big)~ 
\ket{p_1}\otimes\cdots\otimes\ket{p_N} \ ,
\end{split}
\label{eqn:braidedtensor}
\end{equation}
where the vectors $\ket{p}\doteq \check a^\dag(p) \ket{\Omega}\in\hil_1$
are the ordinary Fock space states. This defines a deformation of the
tensor product of single-particle states to the braided (or twisted)
tensor product
\beq
\kket{p_1}_\theta\otimes\cdots\otimes\kket{p_N}_\theta\doteq
\ket{p_1}\otimes_\star\cdots\otimes_\star\ket{p_N} \ .
\label{braidedproddef}\eeq
On $\alg_\theta$, the braided tensor product is the extension of the
star product (\ref{starprod}) to the tensor algebra
$\alg_\theta\otimes\alg_\theta$, so that
\beq
f\otimes_\star g\=\big(\mathcal{F}^{(1)}\tri f\big)\otimes
\big(\twist^{(2)}\tri g\big) \= (f\otimes1)\star(1\otimes g) \ .
\label{braidedprodfg}\eeq
With this definition one has
\beq
(1\otimes_\star f)\star(g\otimes_\star1)=\big(\braid^{(2)}
\tri g\big)\otimes_\star\big(\braid^{(1)}\tri f\big) \ .
\label{braidstar}\eeq
In particular, one has $f\star g=\mu_0(f\otimes_\star g)$ where
$\mu_0:\alg_0\otimes\alg_0\to\alg_0$ is the ordinary (untwisted)
pointwise multiplication on the commutative algebra $\alg_0$. Notice
that the derivation given above uses only the associativity of the
star product. It therefore also applies to quantum field theories on
more general noncommutative spaces, provided that one is able to
define momentum eigenstates (i.e.~free particles).

Consider now the indecomposable twisted state vector in
$\hil_N^\theta$ given by
\begin{equation}
\begin{split}
\ket{p_1}_\theta\otimes\cdots\otimes\ket{p_N}_\theta ~\doteq~&
\check a^\dagger_\theta({p_1})\cdots\check a^\dagger_\theta({p_N})
\ket{\Omega} \\[4pt] \= & \ \check a^\dagger({p_1})~
\e^{\frac{\ii}{2}\,p_1\cdot\theta\cdot\check P}~\cdots~ \check
a^\dagger({p_N})~\e^{\frac{\ii}{2}\,p_N\cdot\theta\cdot\check
  P}\ket{\Omega} \\[4pt]
\= & \ \check a^\dagger({p_1})~\e^{\frac{\ii}{2}\,
p_1\cdot\theta\cdot(p_2+\cdots+p_N)} \\ & \times~\check a^\dag(p_2)~
\e^{\frac\ii2\,p_2\cdot\theta\cdot(p_3+\cdots+p_N)}~\cdots~ \check
a^\dagger({p_{N-1}})~\e^{\frac{\ii}{2}\,p_{N-1}\cdot\theta\cdot p_N}~
\check a^\dagger({p_N})\ket{\Omega} \ .
\end{split}
\label{twistvector}\end{equation}
The overall exponential in (\ref{twistvector}) is inverse to the
exponential defining the braided tensor product in
(\ref{eqn:braidedtensor}). Hence the relation between twisted
multiparticle states and ordinary Fock space vectors is inverse to
(\ref{braidedproddef}), and one has
\beq
\ket{p_1}\otimes\cdots\otimes\ket{p_N}=
\ket{p_1}_\theta\otimes_\star\cdots\otimes_\star\ket{p_N}_\theta \ .
\label{braidedtwisted}\eeq
Since the noncommutative star product may be defined in terms of the
braiding of the tensor product alone, it follows that by using both
braiding and twisting one cancels out the effects of
noncommutativity. This is simply another version of the observations
regarding the commutative dipole field operators given in
Section~\ref{FreeDipole}, and is the basis for the observations
of~\cite{Fiore:2007vg,Tureanu:2006pb}. In particular, both formulas
(\ref{braidedproddef}) and (\ref{braidedtwisted}) agree with the
general conclusions of~\cite{Oeckl2000}, derived using methods of
braided quantum field theory, concerning the equivalences between
correlation functions of the commutative and noncommutative field
theories. Notice that by employing both the braided tensor product
{\em and} the twisted oscillators into the definition of the quantum
field theory, the standard flip operators $\sigma$ generate a
representation of the permutation group on the resulting Fock space.

On the other hand, one could have defined the braided Fock space,
i.e. the Fock space defined in terms of braided tensor products, by
using twisted creation and annihilation operators with the {\em
  opposite} twist. This would be tantamount to the definitions
\beq
\begin{array}{rcl}
\check b_\theta(p) & \doteq & \check
a(p)~\e^{\frac{\ii}{2}\,p\cdot\theta
\cdot\check{P}} \ , \\[4pt]
\check b^\dagger_\theta(p) & \doteq & \check a^\dag(p)~
\e^{-\frac{\ii}{2}\,p\cdot\theta\cdot\check{P}} \ .
\end{array}
\eeq
It would manifestly reproduce all the relations for the Moyal star
product of the corresponding quantum fields. We will see in the
following that the Fock space formalism of this section is the only
one which is compatible with duality transformations of the twisted
quantum fields.

\newsection{Twisted Quantum Fields in Magnetic
  Backgrounds\label{TwistMagnetic}}

In this section we will describe how the previous
considerations generalize to charged scalar fields in a constant
background magnetic field. These models define duality-covariant
noncommutative quantum field
theories~\cite{Langmann:2002cc,Langmann:2003if}. Their infrared and
ultraviolet regimes are indistinguishable due to the duality, a
property which eliminates the pathologies associated to UV/IR mixing
and renders the noncommutative field theory
renormalizable~\cite{Grosse:2004yu}. These models naturally arise when
we consider the behaviour of twisted quantum fields under duality
transformations of the noncommutative spacetime, as will be further
elaborated in the subsequent sections. We will find that the effective
spacetime geometry underlying the twisted quantum fields in these
cases is modified by the magnetic background.

\subsection{Duality Covariant Quantum Fields\label{CovTwist}}

We begin by deforming the algebra $\cD_{\theta}$ of differential
operators defined by (\ref{Dthetarels}) to an algebra $\cD_{\theta,F}$
whose generators $\hat x_i,\hat p^j$ obey the commutation relations
\bea
\big[\hat x_i\,,\,\hat x_j\big] &=& \ii\theta_{ij} \ ,
\nonumber\\[4pt] 
\left[\hat x_i\,,\,\hat p^j\right] &=& \ii\delta_i{}^j \ ,
\nonumber\\[4pt]  \left[ \hat p^i\,,\,\hat p^j\right] &=& -\ii F^{ij}
\ ,
\label{DFthetarels}\eea
where $F^{ij}$ is an additional antisymmetric tensor in $GL(d,\real)$
which plays the role of a constant background magnetic field. The
operators $\hat p^i$ are interpreted as magnetic translation
operators, which translate the coordinate operators $\hat x_i$ in the
standard way and commute with the usual Landau hamiltonian
$\frac12\,\sum_i\,\hat p_i^2$ for the motion of charged particles in
the magnetic background $F^{ij}$. The analog of the algebra morphism
(\ref{algmor}) is somewhat more involved in this case, and was
described originally in~\cite{Seiberg:2000zk}--\cite{Sochichiu:2000kr}
in the context of a background independent formulation of
noncommutative Yang-Mills theory. Its existence is guaranteed by
Darboux's theorem which implies that there is a linear transformation
of operators bringing the commutation relations (\ref{DFthetarels})
into the canonical form (\ref{CCRalgQM}).

Thus there exists constant antisymmetric matrices $\Lambda,\Pi\in
GL(d,\real)$ such that the differential operators
\bea
X_i&=&\hat x_i+\Lambda_{ij}~\hat p^j \ , \nonumber\\[4pt]
P^i&=&\hat p^i+\Pi^{ij}~\hat x_j
\label{XiPidefF}\eea
generate the algebra $\cD_0$. Substituting (\ref{XiPidefF}) into the
canonical commutation relations (\ref{CCRalgQM}) shows that $\Lambda$
and $\Pi$ are determined by the $d\times d$ matrix equations
\bea
2\,\Lambda-\Lambda\,F\,\Lambda&=&\theta \ , \nonumber\\[4pt]
2\,\Pi-\Pi\,\theta\,\Pi&=&F \ , \nonumber\\[4pt]
\Lambda\,\Pi-\Lambda\,F-\theta\,\Pi&=&0 \ .
\label{LambdaPieqsgen}\eea
When $F=0$, the first and third equations give
$\Lambda=\frac12\,\theta$ and $\Pi=0$, while the second one is then an
identity. In this case (\ref{XiPidefF}) reduces to the expected
mapping (\ref{Boppshift}) in the absence of the background
field. Conversely, when $\theta=0$ one has $\Pi=\frac12\,F$ and
$\Lambda=0$, and (\ref{XiPidefF}) is just the standard relationship
between momentum and magnetic translation operators for the
propagation of charged particles in a constant magnetic field. Note
the perfect symmetry between position and momentum variables. This is
the crux of the duality-covariant model~\cite{Langmann:2002cc}. In
this case the dipole momenta $P^i$ yield additional nonlocal field
redefinitions in momentum space in terms of magnetic translations, and
hence lead to a covariance under the UV/IR duality between
noncommutative dipoles and elementary noncommutative fields.

It is easy to see that the equations (\ref{LambdaPieqsgen})
cannot be satisfied when $F=\theta^{-1}$. From the first two equations
one shows that $F=\theta^{-1}$ if and only if $\Lambda=\theta$ and
$\Pi=\theta^{-1}$, but this is inconsistent with the third
equation. The case $F=\theta^{-1}$ is somewhat special and must be
handled separately~\cite{Seiberg:2000zk}--\cite{Sochichiu:2000kr}. We
shall therefore deal first with the generic case where
$F\neq\theta^{-1}$.

Both the appropriate analogs of the Weyl transform and the Drinfeld
twist further require commuting momentum operators which generate
translations in the noncommuting coordinates $\hat x_i$ in the
standard way. The operators
\beq
\tilde P^i\=\Big(\frac1{\id_d-\theta\,\Pi}\Big)^i_{~j}~P^j
\label{DiHPdef}\eeq
satisfy the requisite commutation relations
\beq
\big[\tilde P^i\,,\,\tilde P^j\big]\=0 \qquad \mbox{and} \qquad
\big[\hat x_i\,,\,\tilde P^j\big]\=\ii\delta_i{}^j \ .
\label{Dicommrels}\eeq
Note that $\Pi\neq\theta^{-1}$ by our assumption that
$F\neq\theta^{-1}$. The twist operator defining the standard Moyal
product associated to the noncommutative algebra $\alg_\theta$ in the
magnetic background is then obtained by substituting $P^i$ with
$\tilde P^i$ in (\ref{twistelt}). Thus the required abelian Drinfeld
twist associated to the algebra $\cD_{\theta,F}$ is given by
\beq
\tilde\twist\=\exp\big(-\mbox{$\frac\ii2$}\,
\tilde\theta_{ij}\,P^i\otimes P^j\,\big) \qquad \mbox{with} \quad
\tilde\theta~\doteq~\frac1{\id_d-\Pi\,\theta}~
\theta~\frac1{\id_d-\theta\,\Pi} \ .
\label{twisteltF}\eeq

Consider now a non-relativistic complex scalar field with the mode
expansion
\beq
\check\phi(x) =\int\, \frac{\dd^{d}k}{(2\pi)^d}~
\left(\check a( k) ~\e^{-\ii k\cdot x} + \check b^\dag( k) ~
\e^{\ii k\cdot x}\right) \ ,
\label{modeexpfieldF}\eeq
where the pair of creation and annihilation operators
$(\check a(k),\check a^\dag(k))$ and $(\check b(k),\check b^\dag(k))$
generate two mutually commuting copies of the canonical commutation
relation algebra ($2p^0\to1$ in (\ref{CCRalg})). This realizes the
fields (\ref{modeexpfieldF}) as operator-valued distributions
on the Fock space $\hil_{\check a}\otimes\hil_{\check b}$. The Fourier
momenta $k$ in this expansion are identified as the eigenvalues of the
commuting dipole momentum operators $P^i$, while the coordinates $x$
are the eigenvalues of the dipole position operators $X_i$. Similar
expansions of commutative quantum fields in magnetic backgrounds are
considered in~\cite{Kim:2001wb}.

By using the normalized translation generators (\ref{DiHPdef}) to
shift the parity operator
\beq
\weyl{\delta^d(\hat x-\xi)}\doteq\e^{-\ii\xi\cdot\tilde P}~
\weyl{\delta^d(\hat x)}~\e^{\ii\xi\cdot\tilde P} \ ,
\label{parityshift}\eeq
the definition of the Weyl transform is the same as in
(\ref{WeyltransfMoyal}). This is the definition appropriate to the
Moyal product (\ref{WeylMoyalprod}) defined by (\ref{twisteltF}). To
carry out the analogous calculation to that of
(\ref{eqn:twistedcomponent}), we eliminate the magnetic translation
operators $\hat p^i$ in the definition of the dipole coordinates using
the second equation of (\ref{XiPidefF}). Using the commutation
relations (\ref{Dicommrels}) along with the fact that the matrix
$(\id_d-\Lambda\,\Pi)\,\Lambda$ is antisymmetric, we find
\bea
\phi_p^+(X)&=&\int\,\dd^d\xi~\e^{-\ii p\cdot\xi}~\check a(p)~
\weyl{\delta^d(\xi-\hat x-\Lambda\cdot\hat p)} \nonumber\\[4pt]
&=&\int\,\dd^d\xi~\int\,\frac{\dd^{d}k}{(2\pi)^d}~
\e^{\ii(k-p)\cdot\xi}~\check a(p)~\e^{-\ii k\cdot(\hat x+\Lambda
\cdot\hat p)} \nonumber\\[4pt] &=&\check a(p)~
\e^{-\ii p\cdot((\id_d-\Lambda\,\Pi)\cdot\hat x+\Lambda\,(\id_d-
\theta\,\Pi)\cdot\tilde P)} \nonumber\\[4pt] &=&\check a(p)~
\e^{-\frac\ii2\,p\cdot(\id_d-\Lambda\,\Pi)\,(\Lambda\,\theta\,\Pi)
\cdot p}~\e^{-\ii p\cdot\Lambda\cdot
P}~\e^{-\ii p\cdot(\id_d-\Lambda\,\Pi)\cdot\hat x} \ .
\label{phipXF}\eea

This result suggests the definition of twisted oscillators
\bea
\check a_{\theta,F}(p)&\doteq&
\e^{-\frac\ii2\,p\cdot(\id_d-\Lambda\,\Pi)\,(\Lambda\,\theta\,\Pi)
\cdot p}~\check a(p)~
\e^{-\ii p\cdot\Lambda\cdot\check P} \ ,
\nonumber\\[4pt] \check b_{\theta,F}(p)&\doteq&
\e^{-\frac\ii2\,p\cdot(\id_d-\Lambda\,\Pi)\,(\Lambda\,\theta\,\Pi)
\cdot p}~\check b(p)~
\e^{-\ii p\cdot\Lambda\cdot\check P}
\label{twistedoscF}\eea
and similarly for their hermitean conjugates, where
\beq
\check P^i=\int\,\frac{\dd^dk}{(2\pi)^d}~k^i\,\left(\check a^\dag(k)\,
\check a(k)+\check b^\dag(k)\,\check b(k)\right)
\label{totmomab}\eeq
is the total momentum operator on $\hil_{\check a}\otimes\hil_{\check
  b}$. These operators obey the twisted canonical commutation
relations
\bea
\check a_{\theta,F}(p)\,\check a_{\theta,F}(q)&=&\e^{2\ii p\cdot
\Lambda\cdot q}~\check a_{\theta,F}(q)\,\check a_{\theta,F}(p) \ ,
\nonumber\\[4pt]
\check a_{\theta,F}(p)\,\check a^\dag_{\theta,F}(q)&=&\e^{-2\ii p\cdot
\Lambda\cdot q}~\check a^\dag_{\theta,F}(q)\,\check a_{\theta,F}(p)+
\delta^d(p-q) \ , \nonumber\\[4pt]
\check b_{\theta,F}(p)\,\check b_{\theta,F}(q)&=&\e^{2\ii p\cdot
\Lambda\cdot q}~\check b_{\theta,F}(q)\,\check b_{\theta,F}(p) \ ,
\nonumber\\[4pt]
\check b_{\theta,F}(p)\,\check b^\dag_{\theta,F}(q)&=&\e^{-2\ii p\cdot
\Lambda\cdot q}~\check b^\dag_{\theta,F}(q)\,\check b_{\theta,F}(p)+
\delta^d(p-q) \ , \nonumber\\[4pt]
\check a_{\theta,F}(p)\,\check b_{\theta,F}(q)&=&\e^{2\ii p\cdot
\Lambda\cdot q}~\check b_{\theta,F}(q)\,\check a_{\theta,F}(p) \ ,
\nonumber\\[4pt]
\check a_{\theta,F}(p)\,\check b^\dag_{\theta,F}(q)&=&\e^{-2\ii p\cdot
\Lambda\cdot q}~\check b^\dag_{\theta,F}(q)\,\check a_{\theta,F}(p) \
, \nonumber\\[4pt]
\check a^\dag_{\theta,F}(p)\,\check b^\dag_{\theta,F}(q)&=&\e^{2\ii
  p\cdot\Lambda\cdot q}~\check b^\dag_{\theta,F}(q)\,\check
a^\dag_{\theta,F}(p)
\label{twistedCCRF}\eea
along with their hermitean conjugates. They generate a twisted Fock
space which is no longer a free product of independent Fock
spaces $\hil_{\check a}^{\theta,F}$ and $\hil_{\check
  b}^{\theta,F}$. In particular, the (untwisted) transposition of
particle and antiparticle states produces nonlocal correlations.

The corresponding field operators
\beq
\check\phi^{\theta,F}(x) =\int\, \frac{\dd^{d}k}{(2\pi)^d}~
\left(\check a_{\theta,F}( k) ~\e^{-\ii k\cdot x} + \check
  b_{\theta,F}^\dag( k) ~\e^{\ii k\cdot x}\right)
\label{modeexptwistedfieldF}\eeq
obey the dipole field relation
\beq
\check\phi(X)=\check\phi^{\theta,F}\big((\id_d-\Lambda\,\Pi)
\cdot\hat x\big) \ .
\label{dipolefieldrelF}\eeq
From (\ref{dipolefieldrelF}) we learn that the twisted field operators
together with the original star product defined by the twist element
(\ref{twisteltF}) do \emph{not} generate a commutative algebra, due to
the presence of the magnetic background $\Pi\neq0$. Instead, the
commutative dipole field algebra is obtained from the Moyal product
with a background dependent redefinition of the noncommutativity
parameter $\theta\mapsto\theta^F$ given by
\beq
\theta^F=(\id_d-\Pi\,\Lambda)~\theta~(\id_d-\Lambda\,\Pi) \ .
\label{thetaPi}\eeq
In particular, the twisted spacetime symmetries act on a new
noncommutative algebra $\alg_{\theta^F}$. The new twisting is a
consequence of the appearence of magnetic translation operators $\hat
p^i$ in the second line of (\ref{phipXF}). This is a realization of
the proposal of~\cite{Chaichian:2006wt} for defining twist elements
associated to gauge symmetries using covariant derivatives.

Such a transformation of the moduli of the scalar field theory is
anticipated by the UV/IR duality
transformations~\cite{Langmann:2002cc}, and it is similar to the
Seiberg-Witten transformation of the magnetic field in terms of open
string degrees of freedom~\cite{Seiberg:1999vs}, which relates
commutative and noncommutative descriptions of the scalar field
theory. As we will discuss further in the ensuing sections, such
changes in the noncommutative geometry relate the twisted quantum
field theories through duality transformations obtained by varying the
two-form $F$. Notice as well that the braiding of multiparticle states in
the twisted Fock space $\hil^{\theta,F}$ generated by the quantum
operators (\ref{twistedoscF}) is determined by yet another generically
distinct noncommutativity parameter $2\Lambda$. Thus the duality
covariant field theory provides a dynamical model for the families of
field theories labelled by different noncommutativity parameters in
e.g.~\cite{Doplicher:1994tu,Fiore:2007vg,Grosse:2007vr,Joung:2007qv},
wherein the twistings of the coordinate algebra and of the quantum
field operators are generically different.

Finally, we note that one can employ a more ``covariant'' definition
of the shifted parity operator by using the noncommuting magnetic
translations, obtained by replacing $\tilde P^i$ with $\hat p^i$ in
(\ref{parityshift}). One then computes as above that the braiding of
multiparticle states is determined by the noncommutativity parameter
$2(\id_d-\Lambda\,F)\,\Lambda$, while the star product defining the
commutative algebra of twisted quantum fields is determined by
$(\id_d-F\,\Lambda)\,\theta^F\,(\id_d-\Lambda\,F)$. It would be
interesting to investigate further the general structure of solutions
to the defining matrix equations (\ref{LambdaPieqsgen}) to see if and
under what conditions there exists a solution for which the braiding
coincides with the twisting defined by the original Moyal geometry
given by (\ref{twisteltF}). In any case, one should realize that even
in the commutative case $\theta=0$, the quantum fields above are
nonlocal in momentum space with respect to magnetic
translations~\cite{Kim:2001wb}.

\subsection{Dimensional Reduction of Self-Dual Quantum Fields}

We now consider the special choice of moduli $F=\theta^{-1}$, wherein
there is no algebra morphism between $\cD_{\theta,\theta^{-1}}$ and
$\cD_0$. This is the self-dual point wherein the noncommutative
quantum field theory is \emph{invariant} under the UV/IR
duality~\cite{Langmann:2002cc}. At this point the
field theory becomes an exactly solvable matrix
model~\cite{Langmann:2003if}, having an infinite-dimensional
$U(\infty)$ symmetry acting by symplectic diffeomorphisms of the
spacetime, and in perturbation theory the beta-functions of the
coupling constants vanish to all orders~\cite{Disertori:2006nq}. We
will demonstrate that these special features of the self-dual point
can all be understood from the fact that the effective spacetime
dimension seen by twisted quantum fields is reduced to $\frac d2\doteq
n$.

Generally, the differential operators
\beq
\hat d^i\doteq\hat p^i+\big(\theta^{-1}\big)^{ij}\,\hat x_j
\label{hatddef}\eeq
obey the commutation relations
\beq
\big[\hat d^i\,,\,\hat x_j\big]\=0 \qquad \mbox{and} \qquad
\big[\hat d^i\,,\,\hat d^j\big]\=-\ii\big(F-\theta^{-1}\big)^{ij} \ .
\label{dicommrels}\eeq
At the self-dual point $F=\theta^{-1}$, the operators $\hat d^i$ thus
belong to the center of the algebra $\alg_\theta$, which is just
$\complex$ (the constant functions). Up to an irrelevant constant
shift one thus has $\hat d^i=0$ or
\beq
\hat p^i=-\big(\theta^{-1}\big)^{ij}~\hat x_j \ .
\label{hatpiirred}\eeq
This is simply the unique irreducible representation of the Heisenberg
commutation relations, and there is a reduction
$\cD_{\theta,\theta^{-1}}\cong\alg_\theta$. There are no independent
momentum operators and one must instead exploit the irreducibility of
the representation to build the twisted quantum states of the scalar
field theory. This leads to a corresponding reduction of the Fock
space at the self-dual point.

We can choose a basis of $\real^d$ in which the antisymmetric matrix
$\theta=(\theta_{ij})$ assumes its Jordan canonical form
\beq
\theta = \left(
\begin{array}{ccccc}
&-\theta_1&&&\\
\theta_1&&&&\\
&&\ddots&&\\
&&&&-\theta_n\\
&&&\theta_n&
\end{array}
\right)
\eeq
where $d=2n$ and $\theta_a\neq0$ for $a=1,\dots,n$. In this basis the
algebra $\alg_\theta=\bigoplus_a\,\alg_{\theta_a}(\real^2)$ splits
into $n$ mutually commuting blocks of noncommutative two-planes. Then
the operators
\bea
X_a&=&\hat x_{2a-1} \ , \nonumber\\[4pt]
P^a&=&-\frac1{\theta_a}\,\hat x_{2a}
\label{CCRQMred}\eea
for $a=1,\dots,n$ generate the canonical commutation relation algebra
of $\cD_0(\real^{d/2})$. Lacking a set of independent translation
generators for the noncommutative space, we must use these coordinates
as our set of canonically conjugate variables. Both the dipole
coordinates $X_a$ and momenta $P^a$ are now \emph{local} and
commuting.

The corresponding twisted quantum fields are given by
\beq
\check\phi^{\theta,\theta^{-1}}(X)=\int\, \frac{\dd^{n}k}{(2\pi)^n}~
\left(\check a_{\theta,\theta^{-1}}( k) ~\e^{-\ii k\cdot X} + \check
  b_{\theta,\theta^{-1}}^\dag( k) ~\e^{\ii k\cdot X}\right)
\label{twistedfieldselfdual}\eeq
where the operators $(\check a_{\theta,\theta^{-1}}(k),\check
a_{\theta,\theta^{-1}}^\dag(k))$ and $(\check
b_{\theta,\theta^{-1}}(k),\check b_{\theta,\theta^{-1}}^\dag(k))$
generate two mutually commuting copies of the undeformed canonical
commutation relation algebra in $\frac d2$ dimensions. These fields
behave as ordinary, commutative quantum fields. The effects of
noncommutativity have been absorbed into a dimensional reduction of
the effective spacetime. This is analogous to what happens in a system
of charged particles constrained to lie in the lowest Landau level,
where the phase space is degenerate and the wavefunctions depend
on only half of the position coordinates. Untwisted noncommutative
fields living in the algebra $\alg_\theta$ can be quantized
by other
means~\cite{Grosse:2004yu,Langmann:2002cc,Langmann:2003if}. But due to
the locality of the dipole operators in this degenerate case, the
twisted quantum field theory admits only a local commutative
description in half the spacetime dimension. The action of twisted
spacetime symmetries is truncated to a ${\rm Diff}(\real^{d/2})$
subgroup, but now there is an action of $U(\infty)$ by \emph{inner}
automorphisms of the algebra $\alg_\theta$ generating an action of the
symplectomorphism group of $\real^d$ on untwisted noncommutative
fields~\cite{Lizzi:2001nd}.

\newsection{Twisted Quantum Fields on Rational Noncommutative
  Tori\label{sec:fockspaceinthedualtheory}}

The definition \pref{eqn:braidedtensor} leads to the problem, in the
untwisted case, of statistics of particles in the traditional approach
to noncommutative quantum field theory. We noted in
Section~\ref{SymGroupAction} that, if any phase appearing in the
braiding factor respects the condition
\beq
p_a \cdot\theta\cdot p_b \= 2\pi \,n \qquad \mbox{for}
\quad n\in\mathbb Z \ ,
\eeq
then the braiding resulting from the action of the flip
operator $\sigma$ on the corresponding two-particle state
disappears. In this section we will extend these considerations to
cases where the noncommutative field theory is defined on a torus
fulfilling the more general condition
\begin{equation}
p_a\cdot\theta\cdot p_b \= 2\pi \,h \qquad \mbox{for}
\quad h\in\mathbb Q \ .
\end{equation}
In this case the the braiding phase factor in $d=2$ dimensions is of
the form
\begin{equation}
\exp\big(2\pi \ii\mbox{$\frac{l}{N}$}\,(n_1\, m_2 - m_1\, n_2)
\big) \qquad\textrm{with} \quad l,N\in\mathbb N \ , \ l<N \ .
\end{equation}

Quantum field theory with complex scalar fields that are adjoint
sections of a gauge bundle over a noncommutative torus with rational
dimensionless noncommutativity parameter is equivalent, by gauge Morita
equivalence, to a quantum field theory defined on a commutative dual
torus, with related moduli. In this section we will apply the standard
construction of the dual field
theory~\cite{Ambjorn:2000cs,Guralnik:2001pv,szabo} to twisted quantum
fields on a rational noncommutative torus and confirm that our
definition of the twisted Fock space maps to the usual one of
commutative quantum field theory. We consider only the two-dimensional
case here in order to elucidate the braided construction of
Section~\ref{BraidedFock} in as concise a way as possible. More
general duality transformations, including the higher-dimensional
cases, will be treated in depth in the next section.

\subsection{Duality Transformations of Quantum Fields}

The lattice $\Gamma$ defining the original noncommutative torus is
determined by the period matrix $\Sigma$ and the standard square
lattice of unit spacing defined by the vectors of
$\tilde\Gamma\doteq\mathbb Z^2$. The period matrix may be regarded as
a map
\begin{equation}
\big(\Sigma\,:\,\tilde\Gamma~\longrightarrow~\Gamma\big) \in
GL(2,\mathbb R) \ .
\end{equation}
The nonrelativistic quantum field on the noncommutative torus is given
by the mode expansion 
\begin{equation}
\check\phi(x) = \sum_{p\in\Gamma}\, \check a(p) ~\e^{-2\pi \ii p\cdot
\Sigma^{-1} \cdot x} \ ,
\label{orfield}\end{equation}
where throughout we display only the positive energy particle
components for brevity. This field, as a function on the covering
plane, is periodic with periods defined by $\Sigma$,
\beq
\check\phi(x+\Sigma\cdot v_i) \= \check\phi(x) \qquad \mbox{for}
\quad i\=1,2 \ ,
\eeq
where $v_i$ are the vectors of the canonical basis of $\mathbb
Z^2$. Hence it determines a well-defined single-valued field on the
torus $\mathbb R^2/\Gamma$. We will regard it as a section of the
trivial rank one gauge bundle over the torus.

The associated commutative dual field lives in the adjoint
representation of a $U(N)$ gauge group~\cite{Guralnik:2001pv} and can
be defined as
\begin{equation}
\check\phi\,^\vee(x) = \sum_{p\in\Gamma}\, \check a(p)\,
\otimes\, Q^{\alpha(p)}\,P^{\beta(p)}~\e^{\ii\frac{\pi}{N}\,\alpha(p)\,\beta(p)}~
\e^{-2\pi \ii p\cdot\Sigma^{-1}\cdot x} \ ,
\label{eqn:moritadualfield}
\end{equation}
where
\begin{equation}
P ~\doteq~ \left(\begin{array}{ccccc}0&1&&&0 \\ &0&1&& \\
    &&\ddots&\ddots& \\ &&&\ddots&1 \\ 1&&&&0\end{array}\right)
\qquad \mbox{and} \qquad
Q ~\doteq~  \left(\begin{array}{ccccc}1&&&&0 \\ &\e^{\frac{2\pi
        \ii}{N}}&&& \\ &&\ddots&& \\ 0&&&&\e^{\frac{2\pi \ii(N-1)}{N}}
  \end{array}\right)
\end{equation}
are the $SU(N)$ 't Hooft shift and clock matrices which generate the
$N\times N$ matrix algebra $\mat(N,\complex)$ and obey the
commutation relation
\begin{equation}
P \,Q = Q\, P ~\e^{{2\pi \ii}/{N}} \ .
\label{tHooftalg}\end{equation}
The phase factor $\exp[\ii\frac{\pi}{N}\,\alpha(p)\,\beta(p)] $ is
introduced for convenience. It entails the symmetric ordering of the
matrices $P$ and $Q$. We have also defined the linear functionals
\begin{equation}
\begin{array}{ccl}
\alpha(q) & \doteq & q\cdot A\cdot v_1 \ , \\[4pt]
\beta(q) & \doteq & q\cdot A\cdot v_2
\end{array}
\end{equation}
with $A$ a nondegenerate integral matrix in $\mathbb M(2,\mathbb
Z)$. The duality map $M$ between the fields (\ref{orfield}) and
(\ref{eqn:moritadualfield}) is implemented by a correspondence between
projective modules over dual tori.

Let us consider the products of the dual field
$\check\phi\,^\vee$. Using (\ref{tHooftalg}) they can be written as
\begin{equation}
\begin{split}
\check\phi\,^\vee(x)\,\check\phi\,^\vee(x) \= & \sum_{p,q\in\Gamma}\,
\check a(p)\,\check a(q)\, \otimes\, Q^{\alpha(p)}\,P^{\beta(p)}\,
Q^{\alpha(q)}\,P^{\beta(q)}\\ & \qquad\qquad \times~
\e^{\ii\frac{\pi}{N}\,(\alpha(p)\,\beta(p)+\alpha(q)\,
\beta(q))}~\e^{-2\pi \ii (p+q) \cdot\Sigma^{-1}\cdot x} \\[4pt]
\= & \sum_{p\in\Gamma}\, \e^{-2\pi \ii p\cdot\Sigma^{-1}\cdot
  x+\frac{\ii\pi}{N}\,\alpha(p)\,\beta(p)}\,Q^{\alpha(p)}\,P^{\beta(p)}
\\ & \qquad\qquad \otimes~
\sum_{q\in\Gamma}\,\check a(q)\,\check a({p-q})~
\e^{-\ii\frac{\pi}{N}\,\det({A}) \, q\times p} \ .
\end{split}
\end{equation}
The component of the product at each frequency $p\in\Gamma$ is given
by the sum
\beq
\sum_{q\in\Gamma}\,\check a(q)\,\check a({p-q})~
\e^{-\ii\frac{\pi}{N}\,\det({A}) \, q\times p} \ .
\eeq
If we require the duality morphism $M$ between the fields $\check\phi$
and $\check\phi\,^\vee$ to be compatible with the products, then we have
to {\em shrink} each polarization factor
\begin{equation}
\e^{\frac{\ii\pi}{N}\,\alpha(p)\,\beta(p)}\,Q^{\alpha(p)}\,P^{\beta(p)}~
\dashrightarrow~ 1 \ .
\label{eqn:shrink}
\end{equation}
In this (heuristic) way we obtain the inverse image of the product
\begin{equation}
M^{-1}\big[\check\phi\,^\vee(x)\,\check\phi\,^\vee(x)\big] = 
\sum_{p\in\Gamma}\, \e^{-2\pi \ii p\cdot\Sigma^{-1}\cdot
  x}~\sum_{q\in\Gamma}\,\check a(q)\,\check a({p-q})~
\e^{-\ii\frac{\pi}{N}\,\det(A) \, q\times p} \ .
\label{eqn:minvproduct}
\end{equation}
This is the Moyal product $\check\phi(x)\star\check\phi(x)$ of the
fields on the original torus. The noncommutativity parameter
$\theta_{ij}$ can be read off from \pref{eqn:minvproduct} once we
restore dimensions to the wavenumbers $p,q$, and one has
\beq
\theta_{ij}\=\frac{\det(A\,\Sigma)}{2\pi\, N}\,\epsilon_{ij} \= 
\frac{\det(A)}{N} \ \frac{\textrm{Area}}{2\pi} \ \epsilon_{ij} \ .
\eeq
We thus recover the anticipated rational dimensionless
noncommutativity parameter
\beq
\Theta=\frac{2\pi\,\theta}{\textrm{Area}}
\eeq
on the torus $\real^2/\Gamma$.

The dual field \pref{eqn:moritadualfield} is valued in the group
algebra of $SU(N)$, and it satisfies the periodicity conditions
\begin{equation}
\begin{array}{rcl}
P\,\check\phi\,^\vee(x)\,P^{-1} & = &
\check\phi\,^\vee\big(x+\frac{1}{N}\,(\Sigma\, A)\cdot v_1\big) \ ,
\\[4pt] 
Q^{-1}\,\check\phi\,^\vee(x)\,Q & = &
\check\phi\,^\vee\big(x+\frac{1}{N}\,(\Sigma\, A)\cdot v_2\big) \ .
\end{array}
\label{dualperiodicity}\end{equation}
By redenoting the 't Hooft matrices as
\beq
\begin{array}{ccl}
V_1 & \doteq & P \ , \\[4pt]
V_2 & \doteq & Q^{-1} \ ,
\end{array}
\eeq
we can write the two conditions in (\ref{dualperiodicity}) as the
single one
\begin{equation}
\big(V_1^{\zeta_1}\,V_2^{\zeta_2}\big)\, \check\phi\,^\vee(x) \,
\big(V_1^{\zeta_1}\,V_2^{\zeta_2}\big)^{-1} = \check\phi\,^\vee\big(x+
\mbox{$\frac{1}{N}$}\,(\Sigma\, A)\cdot \zeta\big)
\label{eqn:twistboundcond}
\end{equation}
where $\zeta=\zeta_1~v_1+\zeta_2~v_2$ is a vector in $\mathbb Z^2$
whose components in the canonical basis are $\zeta_i$. This shows that
the operator
\beq
\Xi(\zeta)\doteq V_1^{\zeta_1}\,V_2^{\zeta_2}
\eeq
implements a particular type of translation. From the relation
\beq
\Xi(\zeta'\,)\,\Xi(\zeta) = \Xi(\zeta+\zeta'\,) ~
\e^{-\ii\frac{2\pi}{N}\,\zeta_1\,\zeta'_2}
\eeq
we deduce the commutation relations
\begin{equation}
\Xi(\zeta'\,)\,\Xi(\zeta) = \Xi(\zeta)\,\Xi(\zeta'\,)~
\e^{-\ii\frac{2\pi}{N}\,(\zeta_1\,\zeta'_2-\zeta_1'\,\zeta_2)} \ .
\end{equation}
This shows that the operators $\Xi(\zeta)$ determine a representation
of the group of magnetic translations on the adjoint dual scalar field
$\check\phi\,^\vee$ by displacement $\zeta\in\mathbb Z^2$. Because of
the identities
\beq
V_i^N = \id_N
\eeq
the operators $\Xi(\zeta)$ are periodic of period $N$ in each
direction. Thus they actually represent translations by vectors of
$\mathbb Z_N^2$, which is consistent with the basic periodicity of
the field $\check\phi\,^\vee(x)$.

We will use this periodicity constraint to regard the dual scalar
field as an adjoint section of a rank~$N$ gauge bundle over a dual
torus. For this, we interpret the equation \pref{eqn:twistboundcond}
as a set of twisted boundary conditions for the dual field
$\check\phi\,^\vee(x)$, implying that the field is periodic up to a
gauge transformation on a torus defined by a possibly different
lattice $\Gamma^\vee$. This would then allow us to define the dual
quantum field theory on the torus $\mathbb R^2/\Gamma^\vee$. We thus
require that there exists a period matrix $\Sigma^\vee$ and a $U(N)$
gauge transformation $\Omega_\zeta(x)$ such that
\begin{equation}
\Omega_\zeta(x)\, \check\phi\,^\vee(x) \,
\Omega_\zeta(x)^{-1} = \check\phi\,^\vee\big(x+\Sigma^\vee\cdot\zeta\big)
\label{eqn:twistboundgaugecond}
\end{equation}
for any $\zeta$ in some set of linearly independent integral vectors
$\{\zeta^{(1)},\zeta^{(2)}\}$.

Consistency of the gauge transformations in
\pref{eqn:twistboundgaugecond} implies that they must fulfill the
cocycle conditions
\beq
\Omega_{\zeta'}\big(x+\Sigma^\vee\cdot\zeta\big) 
\circ \Omega_\zeta(x) = \Omega_{\zeta}\big(x+\Sigma^\vee\cdot\zeta'\,
\big) \circ \Omega_{\zeta'}(x)
\eeq
and 
\beq
\Omega_{\zeta}\big(x+\Sigma^\vee\cdot\zeta\big) = \Omega_\zeta(x) \ .
\label{eqn:conscond3}\eeq
We can write $\Omega_\zeta(x)$ in terms of magnetic translation
operators, and requiring unitarity leads to the expression
\begin{equation}
\Omega_\zeta(x) = \e^{\ii\zeta\cdot B \cdot x} \otimes \Xi(\zeta)
\label{eqn:gaugeoperator}
\end{equation}
for some matrix $B\in\mat(2,\zed)$. In the commutative case at hand,
we can drop the abelian phase factor in (\ref{eqn:gaugeoperator}), as
the cocycle conditions are automatically satisfied by a locally
constant gauge transformation. This will not be the case when we will
deal with a noncommutative field theory in the next section, since the
star product will introduce an additional phase and one cannot
disentangle $U(N)$ fluxes into $U(1)$ and $SU(N)$ components.

Having chosen a set of linearly independent vectors
$\{\zeta^{(1)},\zeta^{(2)}\}$ to represent the basis of
non-contractible homology cycles of the dual torus, we obtain the
fundamental domain on the covering plane $\mathbb R^2$ spanned by the
two vectors
\beq
\mbox{$\frac{1}{N}$}\,(\Sigma\, A)\cdot \zeta^{(1)} \qquad \mbox{and}
\qquad \mbox{$\frac{1}{N}$}\,(\Sigma \,A)\cdot \zeta^{(2)} \ .
\eeq
We can restrict to the case in which the twisted periodicity
\pref{eqn:twistboundgaugecond} is the smallest possible, i.e. to the
minimal fundamental domain. This can be translated into a condition on
the allowed vectors $\zeta^{(1)}$. The fundamental magnetic
translations are generated by the operators
\beq
T_i\doteq\Xi\big(\zeta^{(i)}\big)
\eeq
obeying the commutation relations
\beq
T_i\,T_j=T_j\,T_i ~\e^{\frac{2\pi \ii}{N}\, Q_{ij}} \ , 
\eeq
where the integral matrix
\beq
Q_{ij}\=-\zeta^{(i)}\times\zeta^{(j)}\=-\frac{N}{\Theta}\;
\frac{\det\Sigma^\vee}{\det\Sigma}\ \epsilon_{ij}
\eeq
represents background flux through the dual torus. The standard
choice of parameters~\cite{Guralnik:2001pv} is given by
\beq
A\=\left(\begin{array}{cc}-c & 0 \\ 0 & 1 \end{array}\right) \ , \quad
\zeta^{(1)} \= \left(\begin{array}{c} m \\ 0 \end{array}\right) \quad
\mbox{and} \quad \zeta^{(2)} \= \left(\begin{array}{c} 0 \\ 1
  \end{array}\right) \ ,
\eeq
with the condition $\textrm{gcd}(m,N)=1$ which ensures that the
periodicity constraint \pref{eqn:twistboundgaugecond} cannot be
satisfied on a smaller torus. We then obtain for the dimensionless
noncommutativity parameter 
\beq
\Theta= -\frac{c}{N} ~\in~ \mathbb Q
\eeq
and for the background magnetic flux
\beq
Q_{ij} = m \ \epsilon_{ij} \ .
\eeq

\subsection{Dual Fock Modules}

We can now compare the Fock spaces on which the quantum fields
$\check\phi\,^\vee$ and $\check\phi$ are defined. The twisted Fock
space $\hil^\theta$ of the quantum field theory on the original
noncommutative torus is defined in terms of the twisted oscillators
$\check a_\theta(p),\check a_\theta^\dag(p)$ in
\pref{def:twistedoscill} (along with their negative
energy antiparticle counterparts $\check b_\theta(p),\check
b_\theta^\dag(p)$ as in Section~\ref{TwistMagnetic}). Using
\pref{eqn:moritadualfield} the Fock space for the adjoint scalar field
$\check\phi\,^\vee$ is then the module given by
$\hil^\theta\otimes\complex^N$ and can be defined in terms of the
oscillators
\beq
\begin{array}{rcl}
\check d(p)& \doteq & \check
a_\theta(p)\,\otimes\, \weyl{Q^{\alpha(p)}\,P^{\beta(p)} }
\ , \\[4pt]
\check d\,^\dagger(p) & \doteq & \check a_\theta^\dag(p)\,\otimes\,
\weyl{P^{-\beta(p)} \,Q^{-\alpha(p)}} \ .
\end{array}
\label{damapping}\eeq

It is easy to see that these oscillators satisfy the commutation
relations
\beq
\check d(p) \, \check d(q)\=\check d(q) \, \check d(p) \ \, 
\e^{-\frac{2\pi \ii}{N}\,\det(A)\,p\times q}~ \e^{\frac{2\pi \ii}{N}\,
\det(A)\,p\times q}\=\check d(q) \, \check d(p) \ ,
\eeq
where the first phase factor comes from the 't~Hooft algebra
(\ref{tHooftalg}) and the second one from the twisted oscillator
algebra (\ref{twistcommrels}) (with $2p^0\to1$). As expected, the two
factors cancel each other and one recovers the standard untwisted
canonical commutation relation algebra (\ref{CCRalg}) appropriate to
the commutative dual complex scalar field theory. In other words,
there is a \emph{natural} isomorphism of Fock modules
$\hil^\theta\otimes\complex^N\cong\hil$ defined by the mapping
(\ref{damapping}). The phase cancellation follows from the fact that
matrix multiplication with the polarization matrix
$\weyl{Q^{\alpha(p)}\,P^{\beta(p)}}$ implements the braided tensor
product in \pref{eqn:braidedtensor} for the present case. This matrix
provides the additional twist by magnetic translation operators
appropriate to the quantum field theory in the background flux
$Q_{ij}$. The definition (\ref{damapping}) agrees with the general
considerations of Section~\ref{CovTwist} and explicitly implements the
proposal of~\cite{Chaichian:2006wt} for gauge covariant twist
elements. Completely analogous considerations also apply by fully
incorporating the negative energy antiparticle components of the
complex scalar fields following the treatment of
Section~\ref{TwistMagnetic}.

\newsection{Twisted Quantum Fields on Irrational Noncommutative
  Tori\label{DualityIrrational}}

When the dimensionless noncommutativity parameter of the torus is not
rational-valued, the dual quantum field theory cannot be defined on a
commutative torus, since via the procedure of the previous section we
would obtain a parameter of the form $\Theta = \frac{\det A}{N}$. To
extend the derivation of the previous section to this case, we have to
resort to a dual field theory that lives on a noncommutative torus as
well. Then both the original and the dual scalar field theory involve
noncommutative scalar fields in the adjoint representation of
respective unitary groups~\cite{Ambjorn:2000cs,szabo}. In the dual
field theory there is again a nontrivial magnetic background. Our goal
is to relate the moduli of both noncommutative field theories. We will
also give a more rigorous treatment of the duality transformation in
the most general case, which will include the transform $M$ of the
previous section as a special case.

\subsection{Duality Transformations of Quantum Fields}

We firstly need the periodic version of the Weyl transform
\pref{WeyltransfMoyal} appropriate to fields on a $d$-dimensional
torus with period matrix $\Sigma$. With our previous conventions one has
\begin{equation}
\begin{split}
W\big[f(x)\big] \= f(\hat x) \= & \int\dd^d\xi~ f(\xi) ~
\weyl{\delta(\hat x-\xi)} \\[4pt] \= & \int \dd^d\xi ~
f(\xi)~\frac{1}{|\det\Sigma\,|}\,\sum_{p\in\Gamma}\, \e^{2\pi \ii p\cdot
  \Sigma^{-1} \cdot (\hat x -\xi)} \\[4pt]
\= & \int \dd^d\xi ~f(\xi)~\frac{1}{|\det\Sigma\,|}\,\sum_{p\in\Gamma}\, 
\e^{-2\pi \ii p\cdot \Sigma^{-1}\cdot \xi} ~\weyl{\prod_{i=1}^d\,
  Y_i^{p^i}} \\[4pt]
\= & \frac{1}{|\det\Sigma\,|}~\sum_{p\in\Gamma}~\Big(\,\int \dd^d\xi ~
f(\xi) ~\e^{-2\pi \ii p\cdot \Sigma^{-1}\cdot \xi}\,\Big)~
\prod_{i=1}^d\, Y_i^{p^i} ~\prod_{i<j} \,\e^{\ii\pi\, \Theta_{ij}\,
  p^i\, p^j}
\end{split}
\label{def:weyltransf}
\end{equation}
where the plane wave coordinate operators
\beq
Y_i \doteq \e^{2\pi \ii v_i \cdot\Sigma^{-1}\cdot \hat x}
\eeq
obey the commutation relations
\beq
Y_i\,Y_j = Y_j\,Y_i~\e^{-2\pi \ii\Theta_{ij}} 
\eeq
and $\Theta_{ij}$ is the dimensionless noncommutativity parameter
\beq
\Theta \doteq 2\pi~\Sigma^{-1}\, \theta\,\big(\Sigma^{-1}\big)^\top \
.
\eeq
The modular group $SL(d,\mathbb Z)$ of the noncommutative torus acts
via the transformations
\beq
Y_i~\longmapsto~\prod_{k=1}^d\, Y_k^{H_{ik}} \quad \mbox{and}
\quad \Theta ~\longmapsto~ H \,\Theta \,H^\top \qquad \mbox{for}
\quad H\in SL(d,\mathbb Z) \ .
\eeq
This modular invariance can be used to bring the antisymmetric matrix
$\Theta$ into its Jordan canonical form. The Weyl transform
(\ref{def:weyltransf}) takes a function $f(x)$ that actually lives on
the covering space $\mathbb R^d$ and automatically gives an
operator-valued function with periodicity given by the matrix
$\Sigma$, and hence a function on the noncommutative torus deformation
of $\mathbb R^d/\Gamma$.

Secondly, we need the generalization of the magnetic translation
operators to arbitrary dimensions $d$. Consider the operators
\beq
\Xi(\zeta)\doteq\prod_{i=1}^d\, T_i^{\zeta_i}
\eeq
where $T_i$ are the fundamental magnetic translation operators obeying
\beq
T_i\,T_j = T_j\,T_i~\e^{\ii\frac{2 \pi}{N}\,Q_{ij}} \ ,
\eeq
with $Q_{ij}$ an antisymmetric $d\times d$ integral matrix
representing the background magnetic fluxes through the two-cycles of
a dual torus as before. They satisfy the commutation relations of the
group of magnetic translations by $\zeta,\zeta'\in\zed^d$ given by
\beq
\Xi(\zeta)\,\Xi(\zeta'\,)=\Xi(\zeta'\,)\,
\Xi(\zeta)~\e^{\ii\frac{2\pi}{N}\,\zeta \cdot Q \cdot \zeta'} \ .
\eeq
Analogously to the previous section, using $SL(d,\mathbb
Z)$ invariance we can fix a basis $\{\zeta^{(i)}\}_{i=1}^d$ of
$\zed^d$ in which $Q_{ij}$ assumes its normal form
\beq
Q = 
\left(
\begin{array}{ccccc}
&-q_1&&&\\
q_1&&&&\\
&&\ddots&&\\
&&&&-q_n\\
&&&q_n&
\end{array}
\right)
\eeq
where $d=2n$ and $q_1,\dots,q_n$ are integers.

At this point, we use the fact that the $N$-dimensional
representations of the basic operators $T_i=\Xi(\zeta^{(i)})$ can be
decomposed into a tensor product of several $SU(N_a)$ representations
each generating $\mat(N_a,\complex)$ as an associative algebra. The
ranks of the unitary groups are given by the set of integers
\beq
N_a \doteq \frac{N}{\gcd(N,q_a)}
\label{redrank}\eeq
and by the integer $N_0$ defined via
\beq
N = N_0~\prod_{a=1}^n\, N_a \ .
\eeq
The matrix representations are given by
\begin{equation}
\begin{split}
T_{2a-1}~\doteq~ & \id_{N_1}\otimes\cdots\otimes P_{N_a}
\otimes\cdots\otimes\id_{N_n}\otimes\id_{N_0} \ , \\[4pt]
T_{2a}~\doteq~ & \id_{N_1}\otimes\cdots\otimes Q_{N_a}^{q'_a}
\otimes\cdots\otimes\id_{N_n}\otimes\id_{N_0}
\end{split}
\label{def:decomposedmagtranslation}
\end{equation}
with $a=1,\dots,n$, where $P_h$ and $Q_h$ are the $SU(h)$ 't~Hooft
shift and clock matrices, respectively, and we have defined the
reduced fluxes
\begin{equation}
q'_a\doteq \frac{q_a}{\gcd(N,q_a)} \ .
\label{redflux}\end{equation}
This situation parallels the two-dimensional construction of the
previous section, where we required that the fundamental domain of the
covering space was the smallest one possible for the given periodicity
of an adjoint scalar field. Here we see explicitly how the dimension
$N_a$ of each block in the decomposition
\pref{def:decomposedmagtranslation} and the reduced fluxes $q'_a$ are
interwoven, and the requirement of minimality translates into the
dimension of the representation \pref{def:decomposedmagtranslation}.

We can now consider the gauge transformation operators given formally
by the same formulas as in \pref{eqn:gaugeoperator}. In contrast to
the commutative case, however, we cannot drop the abelian factor. We
rewrite the twisted boundary conditions \pref{eqn:twistboundgaugecond}
using star products defined by a dual noncommutativity parameter
$\theta^\vee$, which will be fixed later, to get
\begin{equation}
\Omega_\zeta(x) \star^\vee \check\phi\,^\vee(x) \star^\vee 
\Omega_\zeta(x)^{-1} =
\check\phi\,^\vee\big(x+\Sigma^\vee\cdot\zeta\big) \ .
\label{eqn:twistboundgaugecondstar}
\end{equation}
But now the cocycle conditions (\ref{eqn:conscond3}) and
\begin{equation}
\Omega_{\zeta'}\big(x+\Sigma^\vee\cdot\zeta) \star^\vee 
\Omega_\zeta(x) = \Omega_{\zeta}\big(x+\Sigma^\vee\cdot\zeta'\,\big) 
\star^\vee \Omega_{\zeta'}(x)
\end{equation}
are no longer trivially satisfied. Instead, we obtain the condition
\begin{equation}
\exp\big(\ii\zeta\cdot\big(B\,\Sigma^\vee-(B\,\Sigma^\vee\,)^\top\,
\big)\cdot\zeta'\,\big)\,\exp\big(-\ii\zeta \cdot (B\,\theta^\vee\,
B^\top\,)\cdot\zeta'\,\big) = \exp\big(\ii\mbox{$
\frac{2\pi}{N}$}\,\zeta'\cdot Q\cdot\zeta\big) \ .
\end{equation}
Since \pref{eqn:conscond3} implies
$B\,\Sigma^\vee+(B\,\Sigma^\vee\,)^\top=0$, we obtain the constraint
\beq
Q = -\frac{N}{2\pi}\,\big(2 B\,\Sigma^\vee- B\,\theta^\vee B^\top\,
\big)
\label{Qconstr}\eeq
on the matrix of magnetic fluxes through the two-cycles of the dual
torus.

We have to impose the twisted boundary conditions on the magnetic
background as well. Defining the background abelian gauge field
\beq
A_i(x) \doteq \mbox{$\frac{1}{2}$}\,\Phi_{ij}\,x^j\otimes \id_N
\label{Aixdef}\eeq
with $\Phi_{ij}$ a constant antisymmetric $d\times d$ matrix, one has
the constraint
\beq
{A}_i\big(x+\Sigma^\vee\cdot\zeta\big) = \Omega_\zeta(x)\star^\vee
{A}_i(x)\star^\vee\Omega_\zeta(x)^\dagger - \ii \partial_i
\Omega_\zeta(x)\star^\vee\Omega_\zeta(x)^\dagger \ .
\eeq
This yields the consistency condition
\beq
\Phi \= \frac{2 B^\top}{\Sigma^\vee-\theta^\vee\,B^\top} \qquad
\mbox{and} \qquad B^\top \= \frac{1}{\id_d+\frac{1}{2}\,\Phi\,
\theta^\vee}\,\frac{1}{2}\,\Phi\,\Sigma^\vee \ .
\eeq
Substituting into (\ref{Qconstr}) we obtain 
\beq
\frac{2\pi}{N~\id_d-Q\,\Theta^\vee}\,Q\=\big(\Sigma^\vee\,\big)^\top\,
\left(\Phi+\mbox{$\frac14$}\,\Phi\,\theta^\vee\,\Phi\right)\,
\Sigma^\vee \=-\big(\Sigma^\vee\,\big)^\top\, {F}\,\Sigma^\vee
\eeq
where 
\beq
{F}_{ij}\doteq D_i{A}_j - \partial_j{A}_i
\eeq
is the noncommutative field strength of the background gauge field
$A$. Notice the contribution from noncommutativity coming from the
gauge covariant derivative operators $D_i$ corresponding to the
background (\ref{Aixdef}).

In order to have operator-valued functions fulfilling the constraint
\pref{eqn:twistboundgaugecondstar}, we have to modify the definition
of the Weyl transform \pref{def:weyltransf} for dual fields. We
can do this by introducing factors made out of the fundamental
magnetic translation operators $T_i$, in a similar fashion to what we
did for the rational case of the previous section. Thus we define the
dual Weyl transform
\begin{equation}
W^\vee\big[f(x)\big] = \frac{1}{|\det\Sigma\,|}~
\sum_{p\in\Gamma}\, a_p(f)~\otimes~\prod_{i=1}^d\,T_i^{p\cdot A\cdot
  v_i} ~\otimes~ \prod_{i=1}^d\, Y_i^{p\cdot C\cdot v_i}~ \prod_{i<j}\, 
\e^{\ii\pi\, \Theta^\vee_{ij}\, p^i\, p^j}
\label{eqn:dualfieldnonq}
\end{equation}
where $A$ and $C$ are nondegenerate $d\times d$ matrices with
$A\in\mat(d,\zed)$ as before, and $a_p(f)\in SU(N_0)$ replace
the Fourier coefficients of the expansion of the field $f(x)$ in
\pref{def:weyltransf}. We can define new plane wave coordinate
operators
\beq
\tilde Y_i = \prod_{j=1}^d\,Y_j^{v_i\cdot C\cdot v_j} ~\otimes ~
\prod_{j=1}^d\, T_j^{v_i\cdot A\cdot v_j}
\label{tildeYidef}\eeq
which satisfy the commutation relations
\begin{equation}
\tilde Y_i \star^\vee \tilde Y_j = \tilde Y_j\star^\vee\tilde Y_i~
\e^{2\pi \ii (-C\,\Theta^\vee\, C^\top + A \,Q'\,N'^{-1}\, A^\top\,)_{ij}}
\ ,
\label{eqn:tildeYcomm}
\end{equation}
where
\beq
Q' = 
\left(
\begin{array}{ccccc}
&-q'_1&&&\\
q'_1&&&&\\
&&\ddots&&\\
&&&&-q'_n\\
&&&q'_n&
\end{array}
\right)
\eeq
is the matrix of reduced fluxes (\ref{redflux}) with respect to a
fixed basis of $\zed^d$ and
\beq
N' = 
\left(
\begin{array}{ccccc}
N_1&&&&\\
&N_1&&&\\
&&\ddots&&\\
&&&N_n&\\
&&&&N_n
\end{array}
\right)
\eeq
is the matrix of reduced ranks (\ref{redrank}). One has the $d\times
d$ integral matrix identity
\begin{equation}
L\, N' + A\, Q' = \id_d
\label{eqn:newbezout}
\end{equation}
arising from the B\'ezout identities for the relatively prime integers
$N_a$ and $q_a'$.

The commutation relations \pref{eqn:tildeYcomm} yield an equation for
the noncommutativity parameter of the original torus given by
\begin{equation}
\Theta = C\,\Sigma^{-1}\,\Sigma^\vee\,\Theta^\vee\, 
\big(\Sigma^{-1}\,\Sigma^\vee\,\big)^\top \,C^\top - A\, Q'\,
N'^{-1}\, A^\top \ .
\label{eqn:relththpq}
\end{equation}
The translation generators in the magnetic background are the
derivatives
\beq
\hat D_i ~\doteq~ \hat\partial_i - \ii \big[{A}_i(\hat x)\,,\,-\big]
\=\hat\partial_i - \mbox{$\frac{\ii}{2}$}\,\Phi_{ij}\, 
\big[\hat x^j\,,\,-\big] \ .
\eeq
Substituting into the standard definition for translation derivations
of the noncommutative torus
\beq
\big[\hat D_i\,,\,\tilde Y_j\big] = 2\pi \ii \big(v_i\cdot\Sigma^{-1}
\cdot v_j\big)~\tilde Y_j
\eeq
we can work out the matrix $C$ to be
\beq
C=\Sigma^{-1}\,\left(\id_d+\mbox{$\frac12$}\,\theta^\vee\,\Phi
\right)\,\Sigma \ .
\eeq

Moreover, from the twisted boundary condition
\pref{eqn:twistboundgaugecondstar} and the matrix B\'ezout identity
\pref{eqn:newbezout} we have
\beq
\id_d \= - N' \,C\,\Sigma^{-1}\,\big(\Sigma^\vee-\theta^\vee\,
B^\top\,
\big) \= - N'\, C\,\Sigma^{-1}\,\left(\id_d+\mbox{$\frac12$}\,
\theta^\vee\,\Phi\right)\, \Sigma^\vee
\eeq
up to an integer matrix. This leads to the relationship
\begin{equation}
\Sigma = \Sigma^\vee\, \left(\Theta^\vee\,Q'-N'\,\right)
\label{eqn:dualsigma}
\end{equation}
between the periods of the dual noncommutative tori. Substituting into
the relation \pref{eqn:relththpq} we obtain
\bea
\Theta &=&
\frac{1}{\Sigma}\,\frac{1}{\left(\id_d+\frac12\,\theta^\vee\,\Phi
\right)^2}\,\Sigma^\vee\,\Theta^\vee\,\frac{\big(\Sigma^\vee\,
\big)^\top}{\Sigma^\top}-A\,\frac{Q}{N}\,A^\top\nonumber\\[4pt]
&=&-\frac{1}{\Theta^\vee\,
Q'-N'}\,\frac{\Theta^\vee}{\;N'^{\top}}+\big(L-N'{}^{-1}\,\big)\,A^\top
\ .
\label{Thetadualrel}\eea
Like the period matrix, the dimensionless noncommutativity parameter
$\Theta$ is defined only modulo an integer matrix on a torus, and so
we can drop the matrix $L\,A^\top$ in (\ref{Thetadualrel}) to finally
obtain the relationship 
\begin{equation}
\Theta = -\frac{1}{\Theta^\vee\,Q'-N'}\,\left(\Theta^\vee\,
L^\top-A^\top\right)
\label{eqn:dualtheta}
\end{equation}
between the noncommutativity parameters of the dual noncommutative
tori in terms of an $O(d,d;\zed)$ transformation.

We have thereby defined an algebra $\cD_{\Theta,F}$ generated by
the operators
\beq
\big\{\hat D_i\,,\,\tilde Y_j\big\} \ ,
\eeq
which we have shown to fulfill the relations of the noncommutative
torus with period matrix $\Sigma$ given by \pref{eqn:dualsigma} and with
dimensionless noncommutativity parameter $\Theta$ given by
\pref{eqn:dualtheta}. It corresponds to the original algebra of
observables. This construction also shows that if we use the Weyl
transform with twisted boundary conditions $W^\vee$ to define a field
on a noncommutative torus with a background magnetic flux, then we can
define a dual field theory on a noncommutative torus with related
moduli and with no background flux, i.e. the fields on the latter
torus are single-valued functions. As expected from the considerations
of Section~\ref{TwistMagnetic}, the momentum operators
in the dual field theory are the generators $\hat D_i$ of the magnetic
translation group corresponding to the background (\ref{Aixdef}).

The correspondence between the Weyl transform $W$ given by
\pref{def:weyltransf} and the Weyl transform with magnetic background
in \pref{eqn:dualfieldnonq} explains the heuristic rule
\pref{eqn:shrink} for passing to the field theory without flux. In the
noncommutative field theory defined by \pref{eqn:dualfieldnonq} the
components of the dual quantum field $\check\phi\,^\vee$ at each
frequency $p\in\Gamma$ are given by the tensor product of three
factors
\beq
\check a_p\big(\phi^\vee\,\big)~\otimes~
\Xi(p\cdot A)~\otimes~\weyl{\prod_{i=1}^d\, Y_i^{p\cdot C\cdot v_i}}
\ .
\label{threefactors}\eeq
The quantum field $\check\phi$ in the noncommutative field theory with
trivial background is reobtained from the dual field theory by
identifying $\tilde Y_i$ as the new coordinate operators, and
therefore the components of the field are given by
\beq
\check a_p(\phi)~\otimes~\weyl{\prod_{i=1}^d\,\tilde 
Y_i^{p\cdot C\cdot v_i}} \ .
\label{twofactors}\eeq
Hence by (\ref{tildeYidef}) one has $\check a_p(\phi)=\check
a_p(\phi^\vee\,)$. Thus the noncommutative quantum field theory with
nontrivial magnetic background has the same oscillators as the
noncommutative quantum field theory without the background.

\subsection{Dual Fock Modules}

The dual twisted quantum field theories have different
oscillators, as in the presence of a background flux we must further
twist by the appropriate magnetic translation operators. If the Fock
module $\hil^\theta\otimes\complex^{N_0}\cong\hil^\theta$ of the
original twisted quantum field theory is built on twisted oscillators
$\check a_\theta(p),\check a_\theta^\dag(p)$ satisfying the algebra
(\ref{twistcommrels}) (with $2p^0\to1$), then by \pref{threefactors}
the Fock space of the dual quantum field theory is defined analogously
to the rational case via the oscillators
\begin{equation}
\begin{array}{rcl}
\check d(p) & \doteq & \check a_\theta(p)\, \otimes\,
\weyl{\Xi(p\cdot A)} \ , \\[4pt]
\check d\,^\dagger(p) & \doteq & \check a_\theta^\dagger(p)\, 
\otimes\, \weyl{\Xi(p\cdot A)^\dagger}
\end{array}
\label{dtorusmapping}\end{equation}
acting on $\hil^\theta\otimes\complex^N$ for $p\in\Gamma$. Using
\pref{eqn:relththpq} and \pref{twistcommrels} the algebra of these
dual creation and annihilation operators is given by
\begin{equation}
\begin{array}{rcl}
\check d(p) \,\check d(q) & = &  \check d(q)\, \check d(p) ~
\e^{2\pi\ii p\cdot\Theta^\vee\cdot q} \ , \\[4pt]
\check d\,^\dagger(p) \,\check d\,^\dagger(q) & = & 
\check d\,^\dagger(q) \,\check d\,^\dagger(p) ~
\e^{2\pi\ii p\cdot\Theta^\vee\cdot q} \ , \\[4pt]
\check d(p) \,\check d\,^\dagger(q) & = & \check d\,^\dagger(q) \,
\check d(p) ~\e^{-2\pi\ii p\cdot\Theta^\vee\cdot q} + 
\delta^d(p-q) \otimes\id_N \ .
\end{array}
\label{dtoruscommrels}\end{equation}

These commutation relations obviously reduce to the ones of
Section~\ref{sec:fockspaceinthedualtheory} in the case when
$\Theta^\vee=0$. The dual oscillators thus form a representation of
the twisted canonical commutation relation algebra with
noncommutativity parameter $\theta^\vee$, and hence the mapping
(\ref{dtorusmapping}) provides a natural isomorphism of twisted Fock
modules $\hil^\theta\otimes\complex^N\cong\hil^{\theta^\vee}$. Again
the negative energy antiparticle components of the complex scalar
fields are straightforwardly incorporated into this discussion
following the treatment of Section~\ref{TwistMagnetic}. We conclude
that the twisted Fock module transforms covariantly under Morita
duality, given our definition for the twisted oscillators as above.

Let us cast this conclusion into a more algebraic framework which can
be extended to the description of dual twisted symmetries for more
general noncommutative spacetimes. Let $\alg$ be the noncommutative
torus algebra with noncommutativity parameter $\theta$, and
$\alg^\vee$ its dual with noncommutativity parameter
$\theta^\vee$. The statement that these two algebras are (strongly)
Morita equivalent means that there exists an
$(\alg,\alg^\vee\,)$-bimodule $\mathcal{M}^\vee$ and an
$(\alg^\vee,\alg)$-bimodule $\mathcal{M}$ with the property that for
any left $\alg$-module $\Ecal$, the module
$\Ecal^\vee=\Mcal\otimes_{\alg}\Ecal$ is a left $\alg^\vee$-module
such that $\mathcal{M}^\vee\otimes_{\alg^\vee}\Ecal^\vee=
\Mcal^\vee\otimes_{\alg^\vee}\Mcal\otimes_{\alg}\Ecal=\Ecal$. The above
calculation shows that this correspondence extends to the quantum
level for the field operators acting on the dual Fock modules
$\Ecal\otimes\hil^\theta$ and $\Ecal^\vee\otimes\hil^{\theta^\vee}$,
given our definition for braiding of multiparticle states. The
bimodule property implies that the left and right actions of the
algebras commute, $(f\tri\psi)\triangleleft
f^\vee=f\tri(\psi\triangleleft f^\vee\,)$ for all $f\in\alg$,
$f^\vee\in\alg^\vee$ and $\psi\in\Mcal^\vee$. When tensored with the
Fock bimodule $\hil^\theta\otimes(\hil^{\theta^\vee})^*$, this is the
content of the correspondence between the mode expansion coefficients
(\ref{threefactors}) and (\ref{twofactors}) of dual quantum
fields. This equivalence also leads to a kind of duality between the
Hopf algebras of twisted spacetime symmetries acting on the algebras
$\alg$ and $\alg^\vee$. It would be interesting to understand the
duality more thoroughly at this algebraic level.

\subsection*{Acknowledgments}

We thank F.~Lizzi for helpful discussions. This work was supported in part
by the EU-RTN Network Grant MRTN-CT-2004-005104. The work of M.R. was
supported in part by a Fellowship of the {\sl Fondazione Angelo della
  Riccia}.

\end{document}